\documentclass[useAMS,usenatbib]{mn2e}

\newcommand{\HI}{\textsc{H\,i}}
\newcommand{\kms}{km s$^{-1}$}

\usepackage{graphicx}
\usepackage{aas_macros}
\usepackage{amsmath,amssymb}
\usepackage{url}

\title[KAT-7 Observations of M83]{Neutral hydrogen and magnetic fields in M83 observed with the SKA Pathfinder KAT-7}

\author[G. Heald et al.]{G. Heald$^{1,2,3}$\thanks{E-mail: George.Heald@csiro.au},
W.~J.~G. de~Blok,$^{2,4,3}$
D. Lucero,$^4$
C. Carignan,$^{4,5}$
T. Jarrett,$^4$
\newauthor
E. Elson$^4$, N. Oozeer,$^{6,7,8}$
T.~H. Randriamampandry,$^4$ and L. van Zee$^9$ \\
$^{1}$CSIRO Astronomy and Space Science, 26 Dick Perry Avenue, Kensington WA 6151, Australia\\
$^{2}$ASTRON, the Netherlands Institute for Radio Astronomy, Postbus 2, 7990 AA, Dwingeloo, The Netherlands\\
$^{3}$Kapteyn Astronomical Institute, University of Groningen, PO Box 800, 9700 AV, Groningen, The Netherlands\\
$^{4}$Department of Astronomy, University of Cape Town, Private Bag X3, Rondebosch 7701, South Africa\\
$^{5}$Observatoire d'Astrophysique de l'Universit\'e de Ouagadougou, BP 7021, Ouagadougou 03, Burkina Faso\\
$^{6}$SKA South Africa, The Park, Park Road, Pinelands, Cape Town 7405, South Africa\\
$^{7}$African Institute for Mathematical Sciences, 6-8 Melrose Road, Muizenberg 7945, South Africa\\
$^{8}$Centre for Space Research, North-West University, Potchefstroom 2520, South Africa\\
$^{9}$Astronomy Department, Indiana University, IN 47405, USA
}
\begin{document}

\date{Accepted 2016 July 12. Received 2016 July 4; in original form 2016 May 16}

\pagerange{\pageref{firstpage}--\pageref{lastpage}} \pubyear{2016}

\maketitle

\label{firstpage}

\begin{abstract}
We present new KAT-7 observations of the neutral hydrogen (\HI) spectral line, and polarized radio continuum emission, in the grand design spiral M~83. These observations provide a sensitive probe of the outer disk structure and kinematics, revealing a vast and massive neutral gas distribution that appears to be tightly coupled to the interaction of the galaxy with the environment. We present a new rotation curve extending out to a radius of 50~kpc. Based on our new \HI\ dataset and comparison with multiwavelength data from the literature we consider the impact of mergers on the outer disk and discuss the evolution of M~83. We also study the periphery of the \HI\ distribution and reveal a sharp edge to the gaseous disk that is consistent with photoionization or ram pressure from the intergalactic medium (IGM). The radio continuum emission is not nearly as extended as the \HI\ and is restricted to the main optical disk. Despite the relatively low angular resolution we are able to draw broad conclusions about the large-scale magnetic field topology. We show that the magnetic field of M~83 is similar in form to other nearby star forming galaxies, and suggest that the disk-halo interface may host a large-scale regular magnetic field.
\end{abstract}


\section{Introduction}

The evolution of galaxies is strongly influenced by their connection to the surrounding environment. Whether the dominant effect is in the form of gas accretion \citep{sancisi_etal_2008}, dynamical interactions \citep[e.g.,][]{boselli_gavazzi_2006}, or mergers \citep[e.g.,][]{blanton_moustakas_2009}, it is in the outskirts of galaxies where these evolutionary effects are clearly visible and traceable. Investigation of these effects is key to understanding the mechanisms by which galaxies continue to grow, or cease growing \citep[see for example][]{putman_etal_2012}. 

In the outer parts of galaxies, neutral hydrogen (\HI) has long been recognized as an excellent tracer of the effects listed above. As a necessary ingredient for future star formation, such material also represents an indication for the potential for continued evolution. By contrast, an ingredient of galactic outskirts whose role is not currently understood is the magnetic field content. Magnetic fields associated with the star-forming disks of galaxies are fairly well understood \citep[e.g.,][]{beck_2016}. However, the properties of magnetic fields in the outer parts of galaxies can so far only be inferred from large-scale synchrotron polarization patterns \citep{braun_etal_2010}, or from polarization properties of distant background radio galaxies \citep{bernet_etal_2013}. Whether the magnetic content of galaxy outskirts is important dynamically \citep[][and references therein]{benjamin_2000,elstner_etal_2014,henriksen_irwin_2016} or energetically \citep[cf.][]{beck_2007} remains unclear.

Probing the outer parts of galaxies can be a difficult observational endeavour because the mass distribution is often extended over a large area. Careful observations with very high sensitivity to low surface brightness features are typically required. While great strides in this direction have been made during recent years, future radio telescopes will provide new capabilities to facilitate this line of research, both for \HI\ and magnetism. Amongst the telescopes that are anticipated in the next few years, two that will provide a large instantaneous field of view (FoV) are the Australian Square Kilometre Array Pathfinder \citep[ASKAP;][]{johnston_etal_2008} and the Aperture Tile in Focus \citep[APERTIF;][]{verheijen_etal_2008} upgrade to the Westerbork Synthesis Radio Telescope (WSRT). A different approach to bolster the speed needed to complete a large survey has been adopted by the Karoo Array Telescope \citep[MeerKAT;][]{jonas_2009}. This latter telescope, despite its comparatively small instantaneous FoV, will nevertheless be particularly well suited for high sensitivity (including low column density) observations over large areas on the sky. For this reason, several MeerKAT surveys are planned that typically aim to probe samples of very faint or distant galaxies \citep[see][]{booth_jonas_2012}. The prospect of combining MeerKAT with the Five hundred meter Aperture Spherical radio Telescope \citep[FAST;][]{nan_etal_2011} holds the promise of extremely sensitive observations that may probe the galaxy-IGM interface \citep{carignan_2016}.

One of the several forthcoming MeerKAT surveys that are currently planned is named ``MeerKAT \HI\ Observations of Nearby Galactic Objects: Observing Southern Emitters'' (MHONGOOSE)\footnote{PI W.~J.~G.~de~Blok; see \url{http://mhongoose.astron.nl/}}. The motivation and design of MHONGOOSE is similar in many respects to the recent Westerbork Synthesis Radio Telescope (WSRT) Hydrogen Accretion in LOcal GAlaxieS \citep[HALOGAS;][]{heald_etal_2011} survey. Specifically, MHONGOOSE will perform deep \HI\ observations of thirty (30) nearby galaxies, achieving column density sensitivities of several $\times10^{18}\,\mathrm{atoms\,cm^{-2}}$ at kpc-scale physical resolution. MHONGHOOSE will build on the capability of most existing \HI\ surveys by also providing excellent polarization data along with the spectral line cubes, thus enabling investigation of the magnetic fields along with the gas morphology and kinematics. As the community prepares for data and science results from MHONGOOSE, initial exploratory observations are being performed with the KAT-7 radio telescope \citep[see, e.g.,][]{lucero_etal_2015,hess_etal_2015,carignan_etal_2016}, which is the precursor radio telescope located on the MeerKAT site in South Africa \citep{carignan_etal_2013}. 

M~83 (NGC~5236) is a grand-design spiral galaxy at the center of a loose group \citep{karachentsev_etal_2007}.
M~83 is a prominent example of an interesting class of galaxies showing extended UV (XUV) emission \citep{thilker_etal_2005,thilker_etal_2007}, indicative of active star formation taking place out to several optical radii (nearly $4\times\,r_{25}$).
What fuels this outer disk star formation? Recent \HI\ observations of M83 have been obtained with the Australia Telescope Compact Array \citep[ATCA;][]{park_etal_2001,jarrett_etal_2013} and with the Very Large Array \citep[VLA;][]{walter_etal_2008,deblok_etal_2008,barnes_etal_2014}.
These reveal a very extended \HI\ distribution that also reaches far beyond the main optical disk, and with a very close morphological correlation between sites of recent star formation and high-column density regions in the \HI\ reservoir \citep{bigiel_etal_2010}. With the available data \citet{bigiel_etal_2010} were able to constrain the outer-disk gas depletion time from {\it in situ} star formation to of order 100~Gyr. In this paper we present new \HI\ observations that reveal even more neutral gas mass in the outer disk than has been seen previously, strengthening the conclusion that a vast, nearly untapped gas mass resides outside of the main disk in this galaxy, and may provide a source of fuel to maintain star formation occurring within the central star forming disk.
We summarize some basic properties of M~83 in Table~\ref{table:m83properties}.

This paper is organized as follows. We describe the KAT-7 data collection and reduction procedures in \S\,\ref{section:data}. The data are described in \S\,\ref{section:analysis}, where we present new details about the \HI\ distribution and kinematics, along with their connection to stellar features in the outer disk and to the environment more generally. We also present new conclusions regarding the large-scale magnetic field structure in M~83 (\S\,\ref{section:polarization}). We conclude the paper in \S\,\ref{section:conclusions}.

\begin{table*}
 \centering
 \begin{minipage}{140mm}
  \caption{Properties of M83.}
  \begin{tabular}{@{}lll@{}}
  \hline
   Property &  Value & Reference \\
 \hline
 Hubble type & SAB(s)c & \citet{devaucouleurs_etal_1991}\\
 Distance & 4.79~Mpc ($43.1\arcsec=1\,\mathrm{kpc}$) & \citet{karachentsev_etal_2007}\\
 $D_{25}$ & $11\farcm7$ & \citet{tully_1988}\\
 $M_B$ & -20.94 & \citet{makarov_etal_2014}\\
 SFR (12$\mu$m) & $4.95\pm0.09\,M_\odot\,\mathrm{yr}^{-1}$ & \citet{jarrett_etal_2013,cluver_etal_2014}\\
 SFR (22$\mu$m) & $3.86\pm0.07\,M_\odot\,\mathrm{yr}^{-1}$ & \citet{jarrett_etal_2013,cluver_etal_2014}\\
 \hline
 Total \HI\ mass & $9.0\times10^9\,M_\odot$ & This work\\
 Stellar mass & $2.88\times10^{10}\,M_\odot$ & \citet{jarrett_etal_2013,cluver_etal_2014}\\
 Virial mass & $2\times10^{12}\,M_\odot$ & \citet{tully_2015} \\
 Systemic velocity & $510\,\mathrm{km\,s}^{-1}$ & This work\\
 Maximum rotational velocity & $170\,\mathrm{km\,s}^{-1}$ & This work\\
 \hline
\end{tabular}\label{table:m83properties}
\end{minipage}
\end{table*}

\section{Data and reduction}\label{section:data}

Our KAT-7 observations were designed to make use of the telescope's flexible correlator capabilities \citep[see][]{carignan_etal_2013} in order to study both the neutral gas content and the magnetic fields in M~83. In these commissioning observations, we observed the target galaxy in several sessions, using two different modes (full-polarisation broadband, and spectral line). MeerKAT observations will allow these to be recorded simultaneously. In this Section we describe how the KAT-7 data were obtained and how the data reduction was performed. A summary of the observations is given in Table~\ref{table:kat7setup}.

\begin{table*}
 \centering
 \begin{minipage}{140mm}
  \caption{KAT-7 observations of M83.}
  \begin{tabular}{@{}lllll@{}}
  \hline
   Date/time & Number of & On-source & mode & mosaic? \\
   \ & antennas & time (h) & \ & \ \\
 \hline
 29 July 2013 & 7 & 7.6 & wideband & N \\
 2-3 March 2014 & 6 & 10.7 & line & Y \\
 15-16 April 2014 & 7 & 7.4 & line & Y \\
 16-17 April 2014 & 7 & 7.0 & line & Y \\
 18 April 2014 & 7 & 5.5 & line & Y \\
 20 April 2014 & 7 & 5.5 & line & Y \\
 22 April 2014 & 7 & 3.6 & line & Y \\
 25 April 2014 & 7 & 3.7 & line & Y \\
\hline
\end{tabular}\label{table:kat7setup}
\end{minipage}
\end{table*}

\subsection{Wideband mode}\label{section:wideband}

Our first KAT-7 observations were performed in a wideband, full-polarization correlator mode. These continuum observations utilized the c16n400M4K correlator mode.  This mode consists of a 400~MHz band with 1024 390.625~kHz channels (of which approximately 650 channels, $\sim256$~MHz, are useable). The correlator was tuned to a center frequency of 1328~MHz.  The data provide correlations in full polarization. The data are interleaved with the gain calibrator (PKS~1313-333; see Table~\ref{table:cals}). Additional calibrators for primary flux and bandpass calibration (PKS~1934-638) and polarization (3C138 and 3C286) were also observed for short periods during the observing run.

\begin{table*}
 \centering
 \begin{minipage}{140mm}
  \caption{Calibrators for M83 data sets.}
  \begin{tabular}{@{}llllll@{}}
  \hline
   Calibrator & RA (J2000) & Dec (J2000) & Flux density (Jy) & Reference & Purpose \\
 \hline
 PKS~1313-333 & 13$^h$16$^m$08\fs0 & $-$33$^d$38$^m$59$^s$ & 1.13 & \citet{tingay_etal_2003} & Gain \\
 PKS~1934-638 & 19$^h$39$^m$25\fs0 & $-$63$^d$42$^m$46$^s$ & 14.98 & \citet{tingay_etal_2003} & Bandpass \\
 3C138 & 05$^h$21$^m$09\fs9 & +16$^d$38$^m$22$^s$ & 8.47 & VLA calibrator manual & Polarization \\
 3C286 & 13$^h$31$^m$08\fs3 & +30$^d$30$^m$33$^s$ & 15.00 & VLA calibrator manual & Polarization \\
\hline
\end{tabular}\label{table:cals}
\end{minipage}
\end{table*}

Known unwanted channels and radio frequency interference (RFI) were flagged automatically using a specialized online routine developed by the KAT-7 team. An additional flagging step to eliminate visibilities corresponding to elevation lower than $20^{\circ}$ was also performed. Clipping using the \textsc{casa} \citep{mcmullin_etal_2007} task {\tt tflagdata} was done manually by looking at the cross-polarisation correlations. After successful flagging, the data were calibrated using PKS~J1934-638 for primary calibration, and PKS~1313-333 as the phase calibrator. The cross-hand delays and phase solutions were determined using 3C286 as the polarization calibrator by assuming a zero polarization model. Following calibration, we recovered for 3C286 a polarization fraction of $8.7\pm0.1\%$ and electric vector position angle (EVPA) of $29.9\pm2.5^\circ$. These extracted values agree reasonably well with the expected values at 1372~MHz \citep[$\approx9.3\%$ and $33^\circ$, respectively; see][]{perley_butler_2013}.

Images were created for each Stokes parameter in \textsc{casa} and deconvolved using a {\tt CLEAN} mask defined by sources cataloged brighter than 10~mJy by the NRAO VLA Sky Survey \citep[NVSS;][]{condon_etal_1998}. Spectral line cubes were produced for Stokes parameters $Q$ and $U$, with frequency resolution $1.5625~\mathrm{MHz}$ and synthesized beam size $244\arcsec\times230\arcsec$. Further analysis steps are described in \S\,\ref{section:polarization}.

\subsection{Line mode}

Subsequent observing sessions focused in on the neutral hydrogen line emission. Our KAT-7 \HI\ observations utilized the c1625M4k correlator mode.  This mode gives an instantaneous bandwidth of 6.25 MHz and 4096 channels each 1.525879~kHz in width.  The correlator was tuned to a center frequency of 1422.8~MHz.  We used three mosaic pointings to cover the full angular extent of the \HI\ disk. These pointings are summarized in Table~\ref{table:m83mosaic}. The observations were set up to observe each pointing in turn for approximately 8~minutes each, followed by a gain calibrator (again PKS~1313-333; see Table~\ref{table:cals}). During each session, the gain calibrator was observed $12-14$ times. The average flux density of the gain calibrator, as measured after bootstrapping the flux calibration from PKS~1934-638, is in every session within $8\%$ of the nominal value given in Table~\ref{table:cals}. The KAT-7 line data sets provide all four correlations but here we consider the parallel hands (HH and VV) only.

\begin{table*}
 \centering
 \begin{minipage}{140mm}
  \caption{KAT-7 line-mode mosaic pointings for M83.}
  \begin{tabular}{@{}lll@{}}
  \hline
   Pointing code & RA (J2000) & Dec (J2000) \\
 \hline
 M83-N & 13$^h$37$^m$00\fs9 & $-$29$^d$26$^m$56\fs0 \\
 M83-C & 13$^h$37$^m$00\fs9 & $-$29$^d$51$^m$55\fs5 \\
 M83-S & 13$^h$37$^m$00\fs9 & $-$30$^d$16$^m$56\fs0 \\
\hline
\end{tabular}\label{table:m83mosaic}
\end{minipage}
\end{table*}

Calibration was performed independently for all observing epochs. We used \textsc{casa} version 4.2.2 for the initial steps and then transitioned to \textsc{miriad} for self-calibration and final imaging (both for the continuum and the \HI\ cubes).

Data flagging (e.g. RFI spikes, antenna shadowing) was performed by hand using the CASA task {\tt flagdata}. The primary (flux and bandpass) calibration was achieved using PKS~1934-638, and subsequent gain calibration (sensitivity and phase variation) was determined using PKS~1313-333. Properties of these calibrators are given in Table~\ref{table:cals}. At this point, offline Doppler correction was performed for the line data since it is not implemented in the correlator. Next, visibilities with $(u,v)$ coordinates satisfying the condition $|u|<5.25~m\approx25\lambda$ were removed from the dataset, even for large values of $v$. This is needed due to an issue, possibly related to internally generated radio frequency interference (RFI), that causes such visibilities have an overly high amplitude, resulting in horizontal stripes in the image plane. In our data this preemptive step typically removes about $4-8$ per~cent of the visibilities, and should not affect our ability to recover extended structures (spacings shorter than $25\lambda$ pick up structures larger than $\approx2.3\degr$, which we do not expect in M83). Moreover, the step that limits the largest accessible angular scale in practice is the flagging for antenna shadowing, which impacts angular scales larger than $\approx0.95\degr$. Next, the data are exported to \textsc{miriad} \citep{sault_etal_1995}.

\begin{figure*}
\includegraphics[width=0.475\hsize]{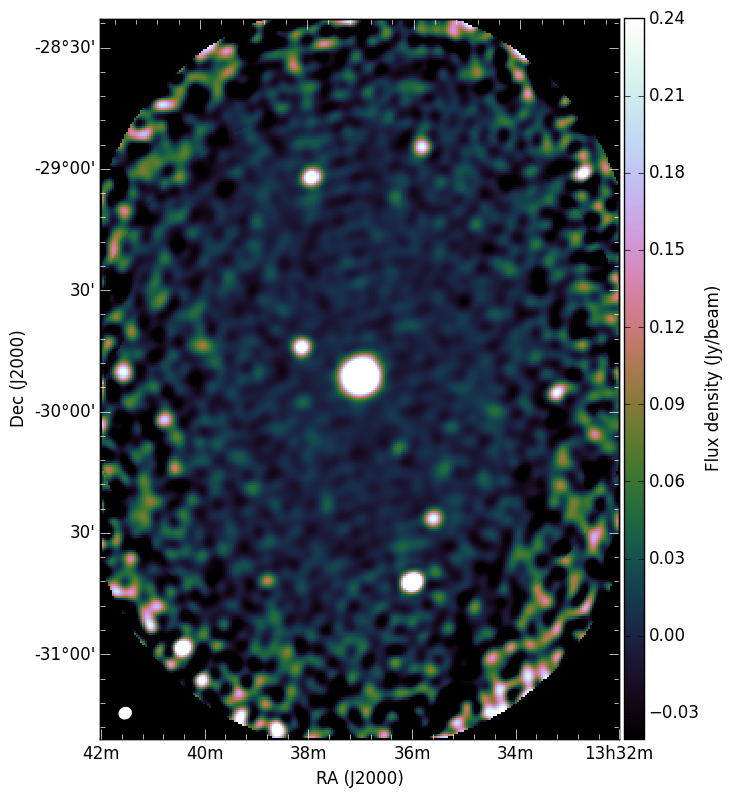}\hfill\includegraphics[width=0.52\hsize]{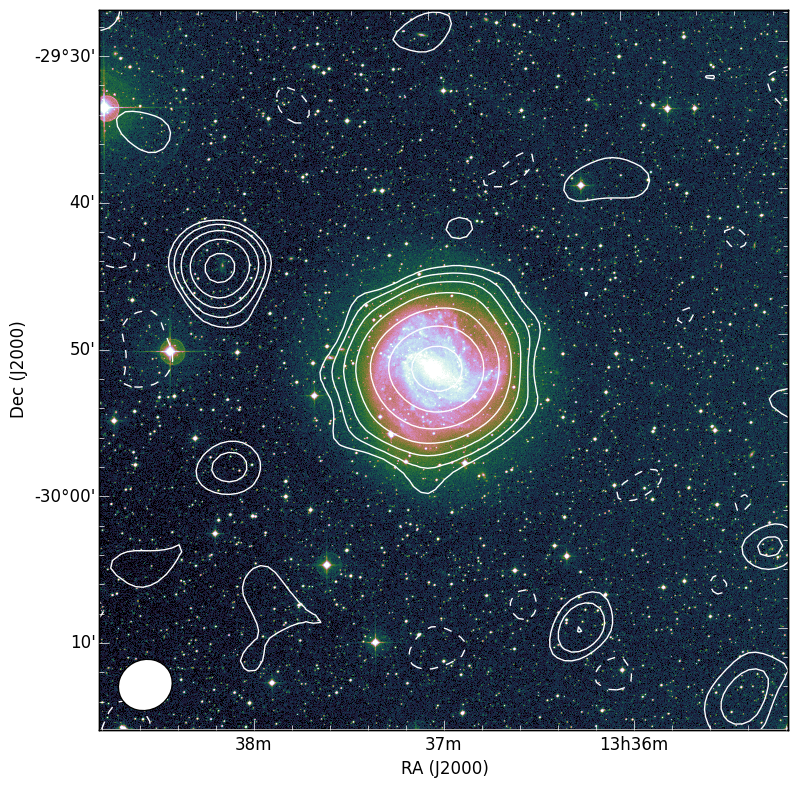}
\caption{Continuum images of the M~83 field, based on the line data. Left: continuum mosaic displayed using the cubehelix colormap. The $225\arcsec\times205\arcsec$ synthesized beam is shown in the bottom left. Right: Optical image (DSS-II red) on a square root stretch, overlaid with contours from the same continuum map. Contours begin at $16\,\mathrm{mJy\,beam}^{-1}$ ($\approx2\sigma$) and increase by powers of two. The $-2\sigma$ level is also displayed with dashed contours. The bright continuum source to the northeast of M~83 is a background galaxy cluster, MCXC~J1338.0-2944, located at $z=0.189$ \citep{piffaretti_etal_2011}.}
\label{figure:continuum}
\end{figure*}

Continuum subtraction was performed in the $uv$ domain using task {\tt uvlin} by fitting the spectrum of each visibility record with a 2nd-order polynomial and subtracting the fit from the data. The corresponding ``channel-0'' (frequency-averaged) visibilities computed from the fitted polynomials were imaged to form a broadband continuum image. That image was deconvolved with the {\tt CLEAN} algorithm, and a model based on the {\tt CLEAN} components was used to update the phase calibration with a timescale of 2~minutes. Subsequently, a single gain amplitude solution was determined per antenna polarization, per 8~minute scan, in each mosaic field. These gain amplitudes were normalized to preserve the flux scale. The resulting high quality set of gain solutions was transferred to the continuum-subtracted line data. The final continuum image was produced using multifrequency synthesis; that image is shown in Figure~\ref{figure:continuum}. 

Finally, the line data were imaged, Hanning smoothed, and deconvolved. We used the task {\tt mossdi} for deconvolving the line cubes. This task performs Steer {\tt CLEAN} \citep{steer_etal_1984}. The deconvolution used a threshold of approximately $0.5\sigma$ and a carefully defined mask (consisting of regions obtained by masking a smoothed version of an earlier cube, rather than boxes). The {\tt CLEAN} mask was derived over several iterations of the \HI\ cube, each with a progressively improved deconvolution and accompanying mask quality.

In order to successfully perform the mosaicing for the line data, the primary beam needs to be specified. We estimated the beam to have a Gaussian profile with a half power beam width (HPBW) given by
\begin{equation}
\mathrm{HPBW}=\alpha\frac{\lambda}{D}
\end{equation}
where $\lambda$ is the observing wavelength, $D$ is the dish diameter, and $\alpha$ is a factor related to the tapering of the dish. The latter is found empirically to have a value of $\alpha=1.27$ for KAT-7, based on holographic measurements \citep{foley_etal_2016}. Along with $\lambda=0.21\,\mathrm{m}$ and $D=12\,\mathrm{m}$ we obtained $\mathrm{HPBW}=4580\arcsec=1.27\degr$. The mosaicing was performed in practice using \textsc{miriad} tasks {\tt invert} with option {\tt mosaic} for initial imaging and the deconvolution task {\tt mossdi} as described above. To perform the final primary beam correction, we used the task {\tt mossen} to determine the effective combined sensitivity map and {\tt maths} to apply it to the cubes.

\begin{figure}
\includegraphics[width=\hsize]{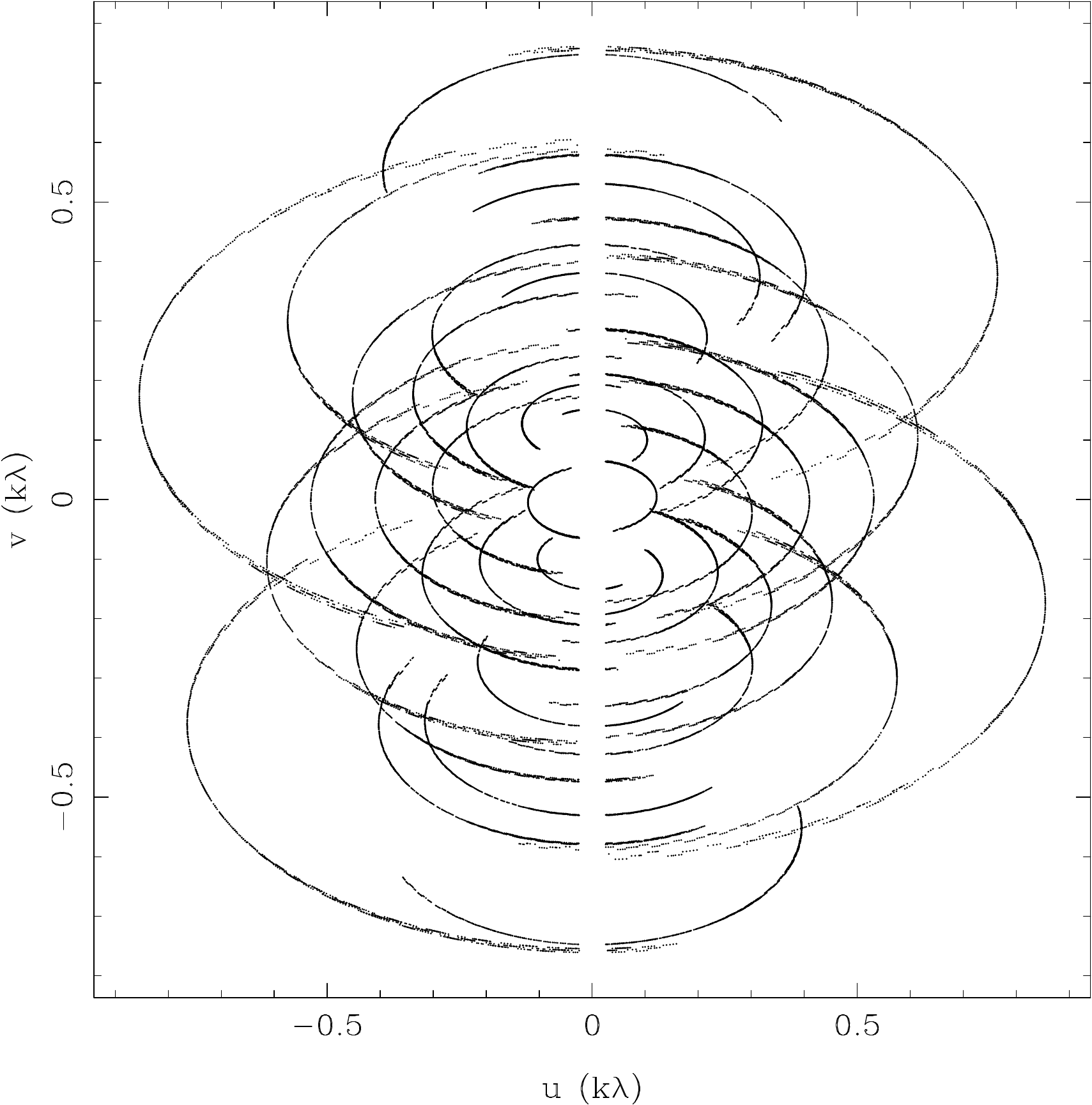}
\caption{uv coverage of the M83 observation. The three mosaic pointings are clearly distinguishable on the longer baselines. Only every eleventh point is plotted.}
\label{figure:uvcov}
\end{figure}

\begin{figure*}
\includegraphics[width=\hsize]{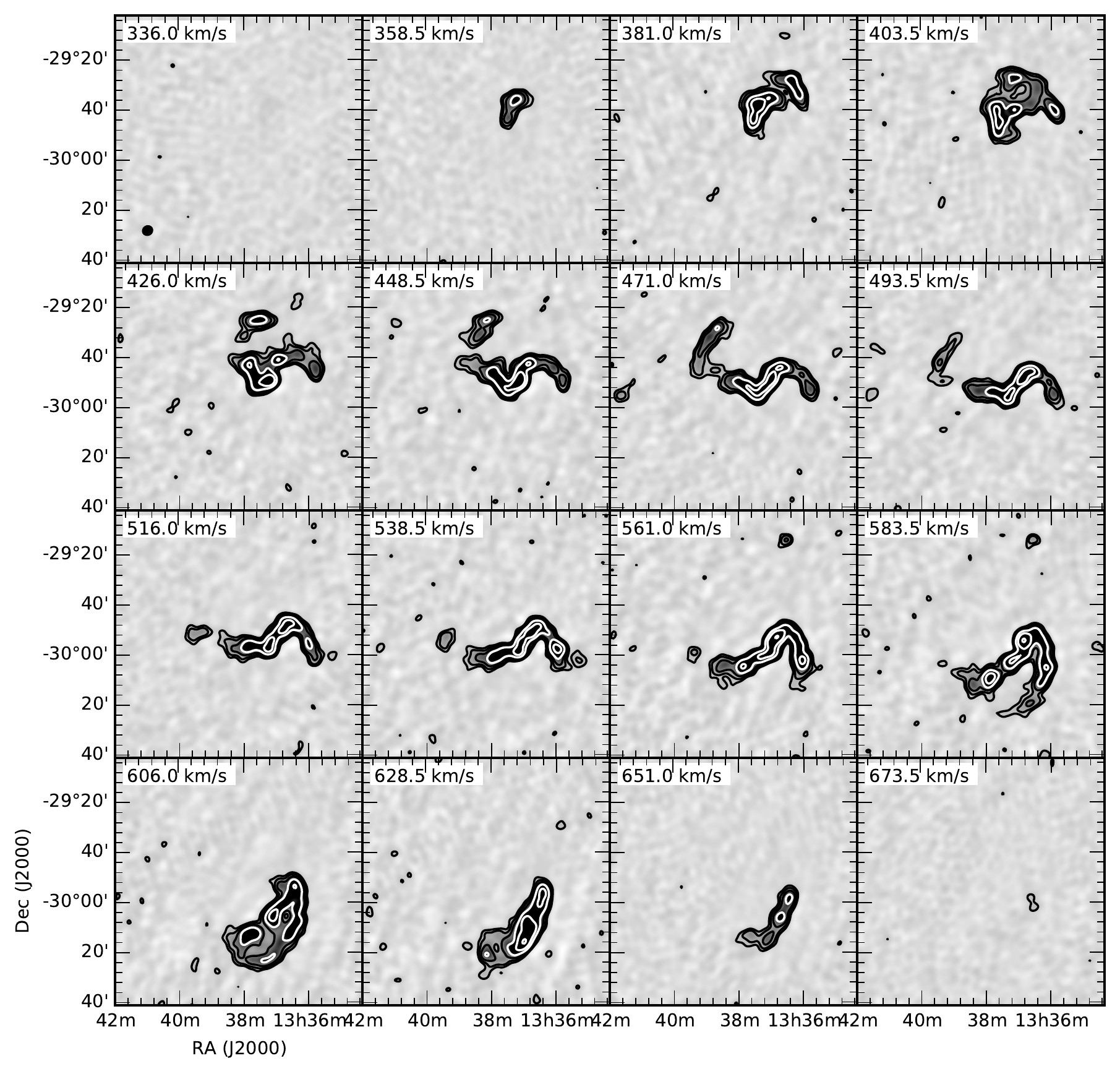}
\caption{Channel maps. Levels start at 25 mJy/beam and increase by powers of two. The velocity of the shown channels is displayed in the upper left of each panel. Only every fifteenth 1.5 km/s channel is shown. The synthesized beam size is shown with a black ellipse in the bottom left corner of the top left frame.}
\label{figure:chanmaps}
\end{figure*}

The $uv$ coverage of the observations is shown in Fig.~\ref{figure:uvcov}. The effect of the $u<25\lambda$ flagging step is plainly visible. Note that while the $uv$ plane is well filled by our tracks, due to the layout of the array there is no strong overdensity of points at the shorter spacings. We found that differences in image-plane resolution and image noise were minimal for a wide range of visibility weighting \citep[robust; see][]{briggs_1995} parameters. Thus we use a single cube for the analysis in this paper, produced using a robust parameter of $+0.5$ and a channel width of $1.5\,\mathrm{km\,s}^{-1}$. The synthesized beam size is $225\arcsec\times205\arcsec$. The noise level per channels is typically $5.0\,\mathrm{mJy\,beam}^{-1}$, and this corresponds to a $3\sigma$ column density sensitivity of $N_\mathrm{HI}=5.6\times10^{18}\,\mathrm{cm}^{-2}$ for a $16\,\mathrm{km\,s}^{-1}$ line width. Channel maps from the \HI\ cube are shown in Figure~\ref{figure:chanmaps}. These properties are summarized in Table~\ref{table:cubeprops}.

\begin{table*}
 \centering
 \begin{minipage}{140mm}
  \caption{M83 \HI\ data cube properties.}
  \begin{tabular}{@{}ll@{}}
  \hline
 Robust parameter & $+0.5$ \\
 Channel width (km/s) & 1.5 \\
 Beam size (arcsec) & $225\times205$ \\
 Beam angle (degr) & $-61$ \\
 Noise level (mJy/beam)\footnote{Per $1.5\,\mathrm{km\,s^{-1}}$ channel.} & 5.0 \\
 $N_\mathrm{HI}$ sensitivity\footnote{Corresponding to $3\sigma$ over a summed line width of $16\,\mathrm{km\,s}^{-1}$.} (atoms\,cm$^{-2}$) & $5.6\times10^{18}$ \\
\hline
\end{tabular}\label{table:cubeprops}
\end{minipage}
\end{table*}

\subsection{Multiwavelength supplementary data}

\subsubsection{GALEX}

We make use of deep GALEX images extracted from the archive. The deep FUV and NUV images were obtained through program GI3\_050007 and included 14256.4~sec in NUV and 14250.4~sec in FUV observed between 2007 March 15 and 2008 March 16. To enhance extended emission associated with star formation in the outer parts of M83, we smooth both FUV and NUV images to $10\arcsec$ resolution.

\subsubsection{WISE}

WISE imaging has proven to be a very effective tool to study the star-formation history of nearby galaxies (Jarrett et al. 2013), owing in part to the (arbitrarily) large field of view and photometric bands that cover both the stellar (3.4 and 4.6 $\mu$m, or the W1 and W2 bands) and gas/dust components (12 and 22 $\mu$m, or W3 and W4 bands). To study M83's disk and greater environment, we constructed dedicated WISE mosaics using a `drizzle' sampling technique \citep{jarrett_etal_2012} that conserves the native angular resolution while providing full coverage of the field. The resulting WISE mosaics cover the galaxy and its dynamic environment, over 4 square degrees.

Global measurements of the star-forming disk were carried out using the characterization pipeline of the WISE Enhanced Resolution Atlas \citep{jarrett_etal_2013}. Foreground stars were identified and PSF-subtracted, the disk shape of M83 was determined and surface brightness was extracted.
The 3.4 $\mu$m (W1) one-$\sigma$ (23.07 mag per sq. arcsec in Vega, or 25.82 mag per sq. arcsec in AB) isophotal diameter is $18.7\arcmin$ with an axis ratio of 0.938.
The corresponding integrated flux densities are 6.24, 3.79, 21.65 and 43.66 Jy, for W1, W2, W3 and W4 respectively.

Demonstrated in \citet{jarrett_etal_2013} and updated in \citet{cluver_etal_2014}, the stellar mass-to-light ratio has a simple dependence on the [W1-W2] color. Assuming our adopted distance of 4.79~Mpc to M83, the corresponding W1 absolute magnitude and in-band luminosity $L_\mathrm{W1}$ are $-24.16$ and $9.120\times10^{10}\,L_\odot$, respectively. The W1-W2 color is 0.12 $\pm$ 0.02 mag, and employing the nearby-galaxy M/L relation from \citet{cluver_etal_2014}, the inferred stellar mass is then $3.020\pm0.390\times10^{10}\,M_\odot$, representing the evolved stellar population that corresponds to the past SF history of M83.

The obscured star formation activity may be estimated by the WISE 12 and 22 $\mu$m luminosities, $\nu L_\nu$, which are $4.07\times10^{9}$ and $4.17\times10^{9}\,L_\odot$, respectively. Correspondingly, we apply the relations in \citet{cluver_etal_2014} to infer the dust-obscured star formation rate (SFR): 5.2 and 4.1 $M_\odot\,\mathrm{yr}^{-1}$, respectively based on the 11.3 $\mu$m PAH emission and the 22 $\mu$m dust continuum. These new SFRs are comparable (within the uncertainties) to the total UV+IR value given by \citet{jarrett_etal_2013}, namely 3.2 $M_\odot\,\mathrm{yr}^{-1}$. In the Local Volume among disk galaxies, the M83 global SFR is significant, within a factor of $\sim$2 of the notable starbursts M82 and NGC~253 \citep{lucero_etal_2015}.

Finally, normalizing the SFR with the stellar mass, the resulting specific SFR is $1.74\times10^{-10}$ and $1.35\times10^{-10}\,\mathrm{yr}^{-1}$, derived from W3 and W4 respectively. Indicative of the large SFR, this disk building rate is relatively high for local galaxies, comparable to other large star forming spirals, including NGC~253 and NGC~6946 \citep{jarrett_etal_2013}.

\section{Neutral gas distribution}\label{section:analysis}

\subsection{General properties}\label{subsection:properties}

\begin{figure}
\includegraphics[width=\hsize]{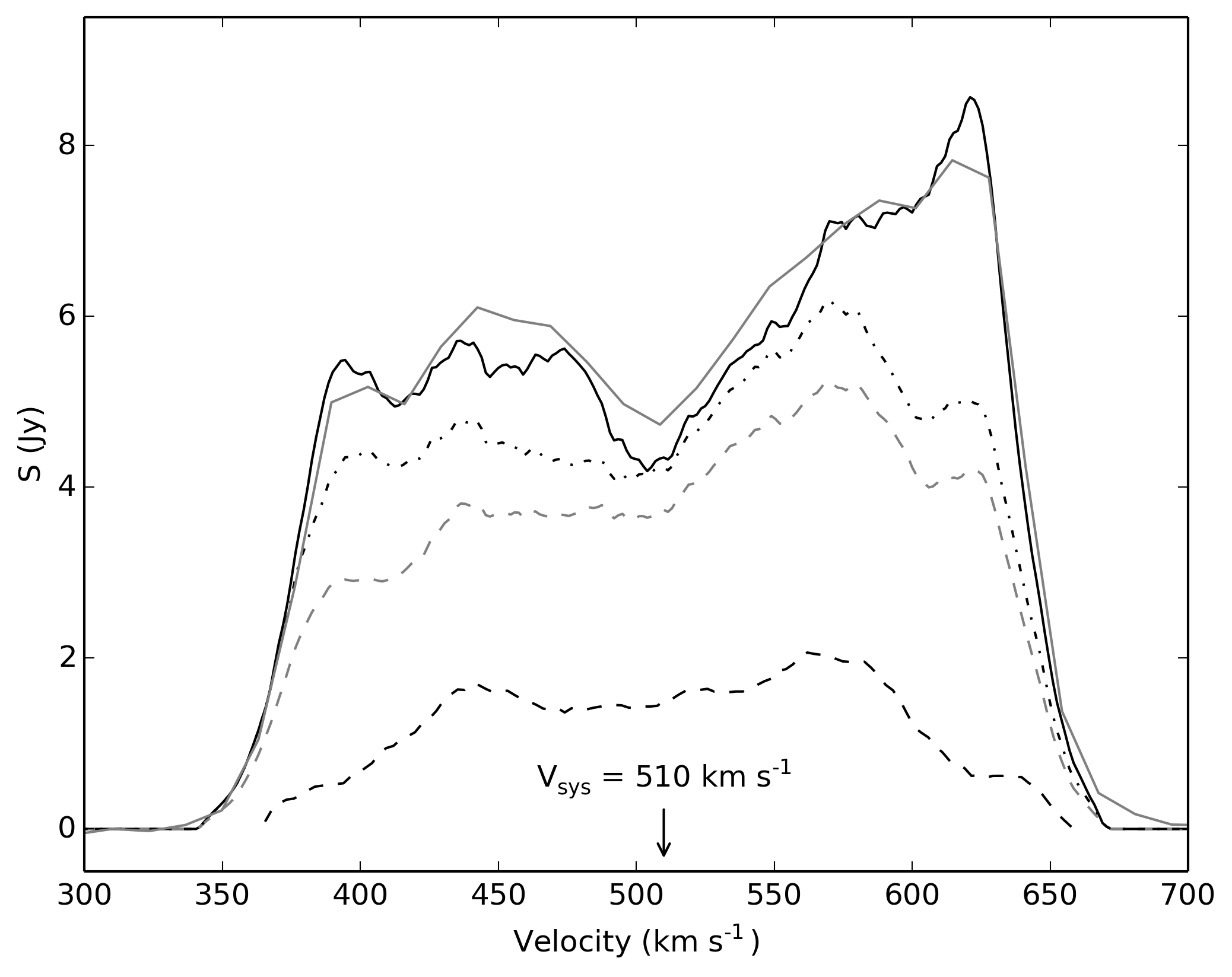}
\caption{Global \HI\ profile of M83 from our KAT-7 observations ({\it solid black line}). The profile from THINGS is also shown for comparison ({\it dashed black line}); here, only about 20\% of the KAT-7 flux density is recovered. The global profile derived by only including \HI\ emission from the KAT-7 data within the 20\% response of the VLA primary beam is also shown for reference ({\it dot-dashed line}), as is the global profile derived by masking the KAT-7 cube using the THINGS moment-0 map ({\it dashed gray line}). Finally, the HIPASS global profile from \citet{koribalski_etal_2004} is also displayed ({\it solid gray line}). The systemic velocity appropriate for the center of the galaxy is indicated with an arrow.}
\label{figure:globprof}
\end{figure}

\begin{figure}
\includegraphics[width=\hsize]{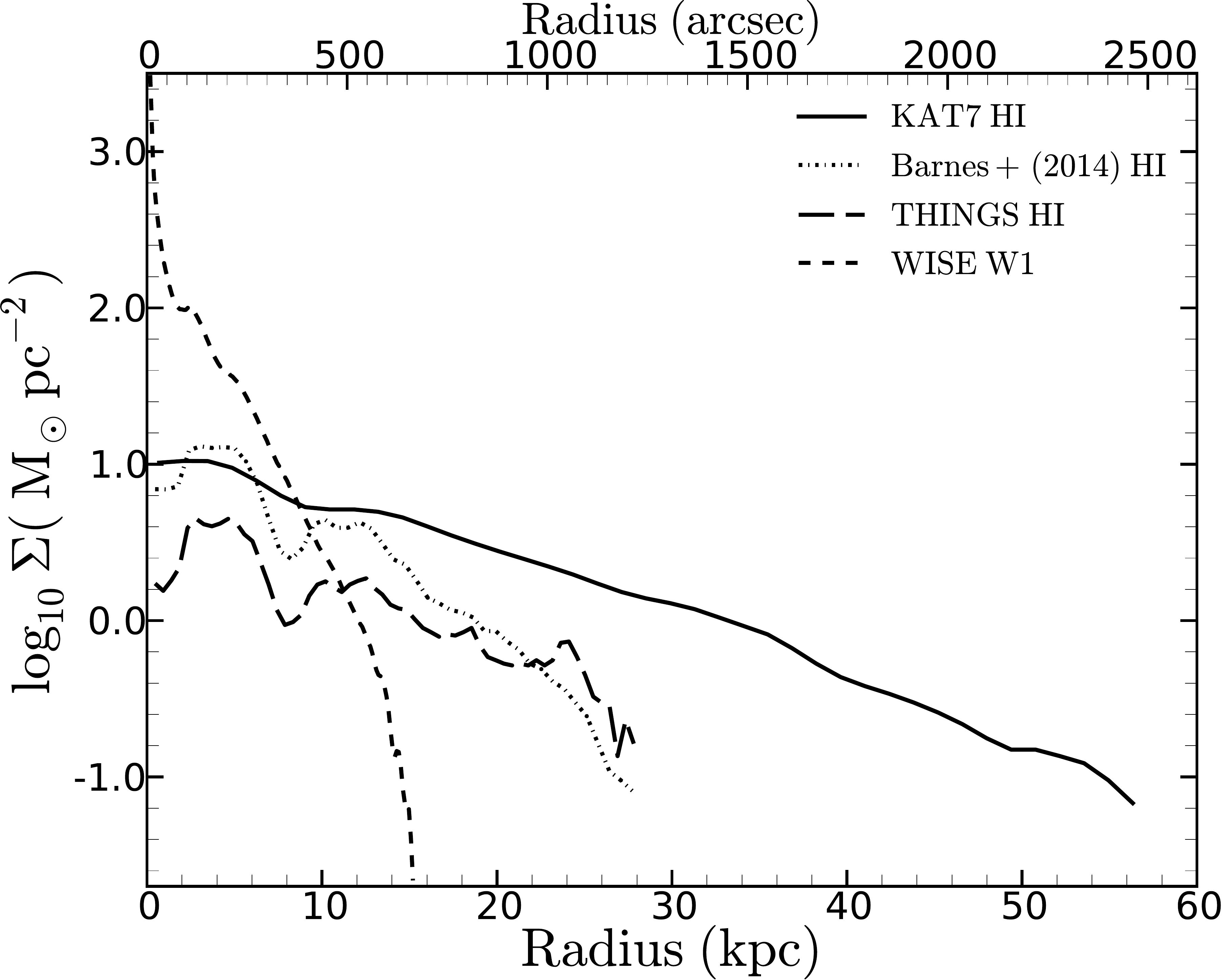}
\caption{Comparison of the \HI\ radial surface density profiles between the KAT-7 data and the VLA data from \citet{deblok_etal_2008} and \citet{barnes_etal_2014}. Also shown is the WISE W1 radio profile.}
\label{fig:sd}
\end{figure}

Based on our KAT-7 \HI\ data cube, we construct a global velocity profile as shown in Figure~\ref{figure:globprof}. The distribution is clearly asymmetric, as has been observed previously. We find that 45\% of the total \HI\ mass is on the approaching side, and 55\% on the receding side. Figure \ref{fig:sd} compares the \HI\ radial surface density profile obtained with KAT-7 with those derived from the VLA data of \citet{deblok_etal_2008} and \citet{barnes_etal_2014}. The flux recovered by KAT-7 on large scales is clearly recognized.

\begin{figure*}
\includegraphics[width=0.43\hsize]{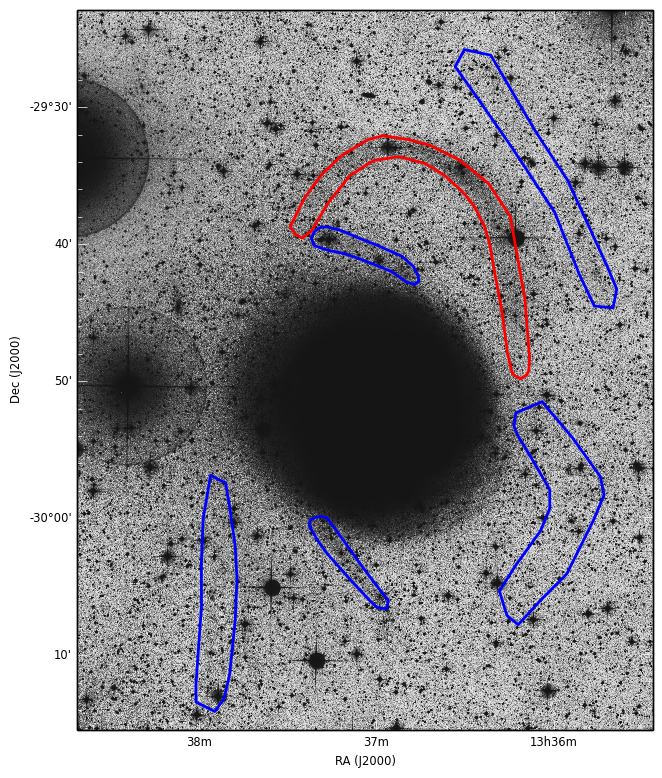}\hfill\includegraphics[width=0.56\hsize]{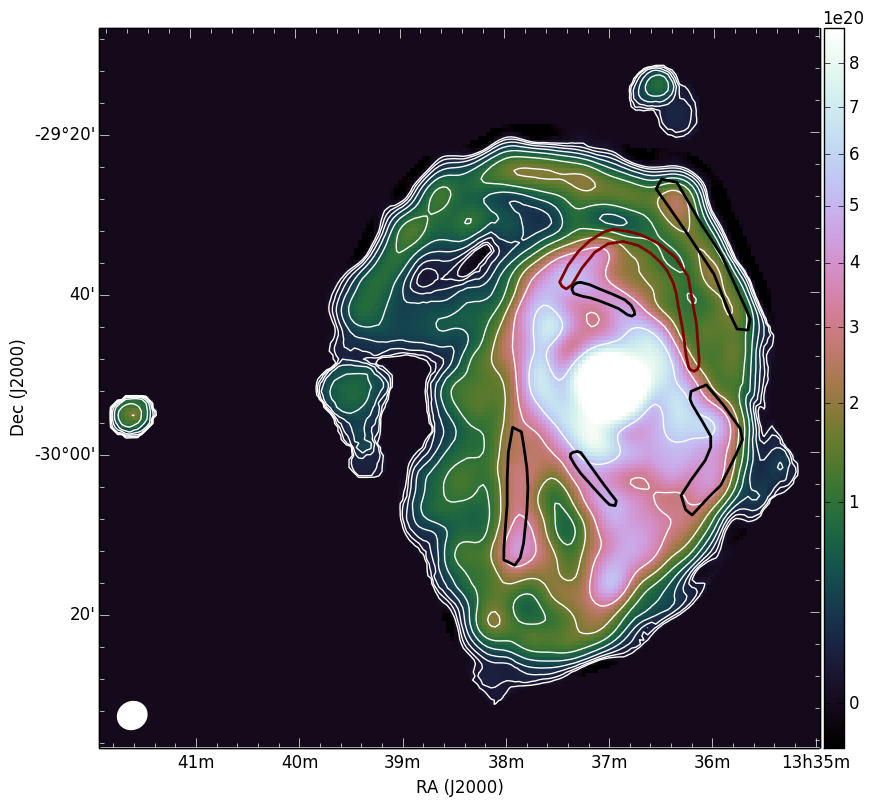}
\caption{Left: Deep image derived from plates taken with the UK Schmidt telescope, courtesy of David Malin \citep[see][]{malin_hadley_1997}, with stellar stream highlighted in red and outer disk star-forming arms highlighted in blue. Right: Neutral hydrogen  column density distribution. Contours start at $5.6\times10^{18}\,\mathrm{cm}^{-2}$ and increase by powers of 1.778. Highlighted regions are the same as in the left panel. The beam size is displayed with a white ellipse in the bottom-left corner of the panel.}
\label{figure:coldens}
\end{figure*}

\begin{figure}
\includegraphics[width=\hsize]{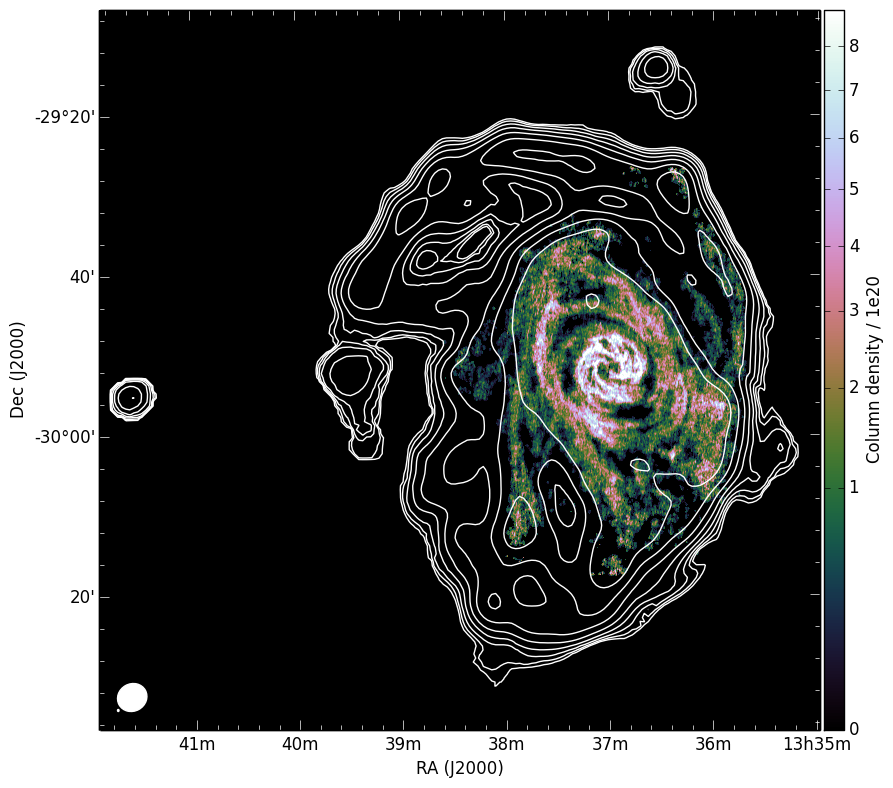}
\caption{Naturally-weighted \HI\ column density map of M83 from THINGS \citep{walter_etal_2008} data, as used by \citet{bigiel_etal_2010}. The KAT-7 contours are overlaid, and demonstrate an excellent morphological agreement despite the vastly different angular resolution, as well as the presence of far more neutral gas in the outer parts in our new observations than has previously been studied. The contour levels are the same as in Fig.~\ref{figure:coldens}. The beam sizes are shown in the bottom-left corner for the KAT-7 data ({\it large white ellipse}) and the VLA data ({\it small white ellipse}).}
\label{figure:thingsoverlay}
\end{figure}

\begin{figure*}
\centering
\includegraphics[width=0.9\hsize]{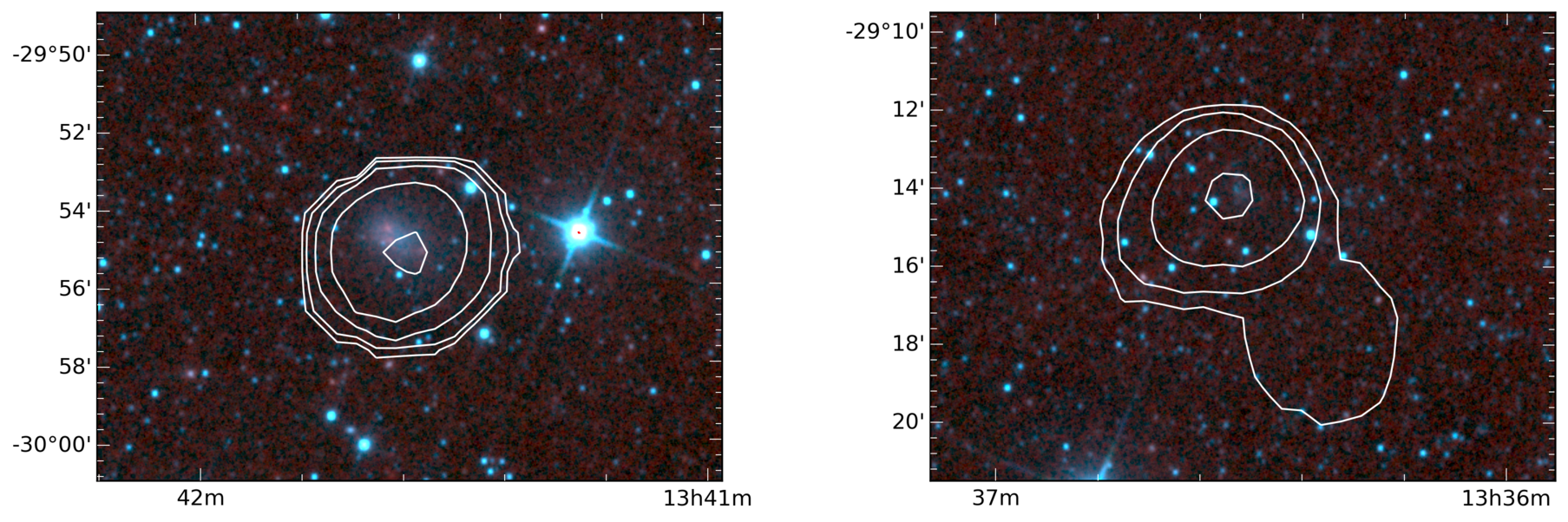}\\
\includegraphics[width=\hsize]{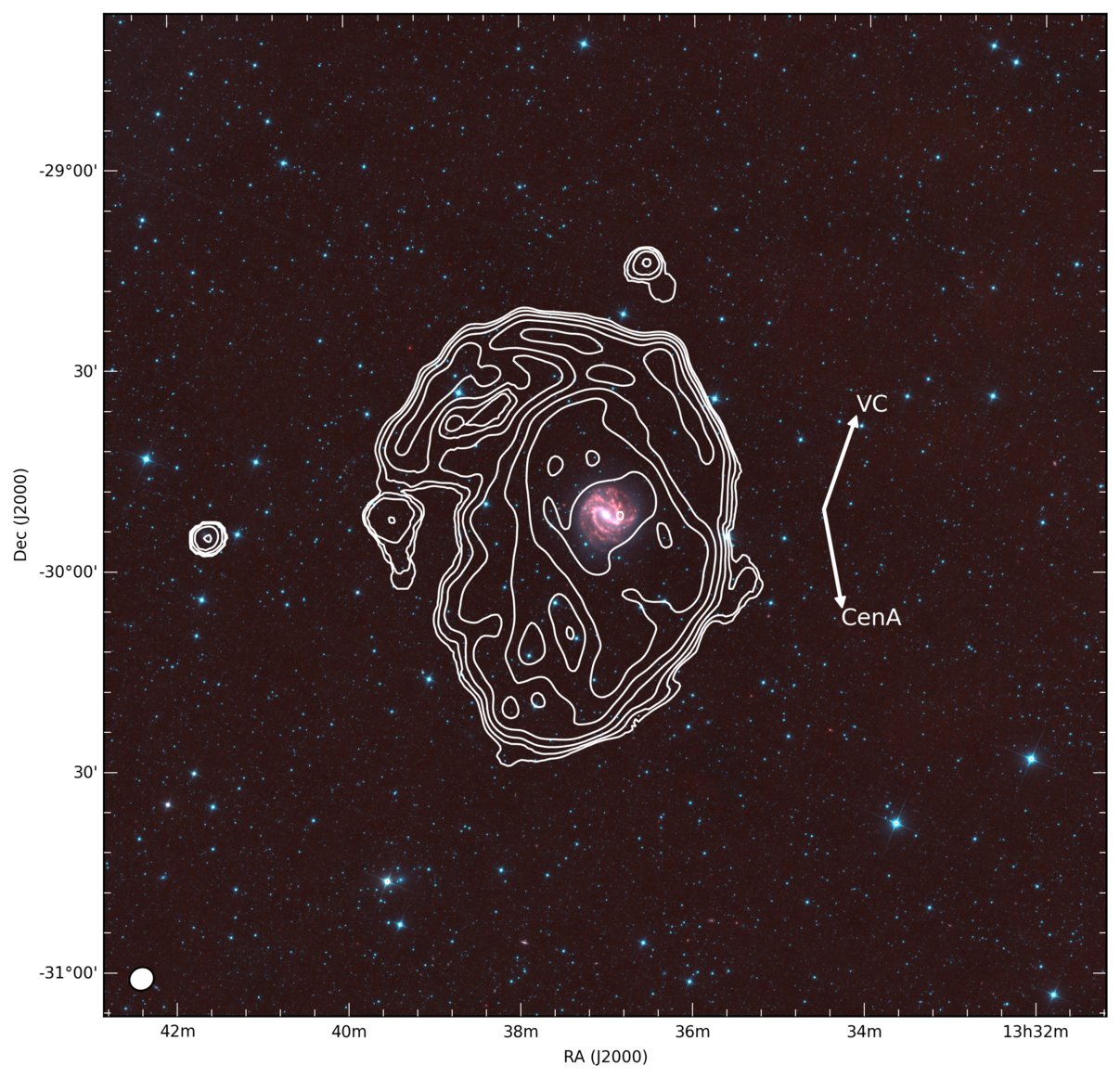}
\caption{\HI\ column density contours (same levels as in Figure~\ref{figure:coldens}), overlaid on the three-color WISE image. In this image, the blue channel is formed from W1 and W2, the green channel from W1, W2, and W3, and the red channel from W3 and W4. Each channel is displayed using a square root stretch. The two arrows indicate the directions toward the Virgo Cluster (VC) and the Centaurus A (CenA) group. Note that UGCA~365 is visible to the north of the extended disk of M~83 itself, and NGC~5264 to the east. These companions are show in greater detail in the upper right and upper left panels, respectively.}
\label{figure:optoverlay}
\end{figure*}

The total mass of \HI\ from our KAT-7 observations is $M_\mathrm{HI}=9.0\times10^9\,M_\odot$, assuming a distance of 4.79~Mpc \citep{karachentsev_etal_2007}.
From the literature, the single-dish \HI\ mass is reported to be $8.8\times10^9\,M_\odot$ \citep{huchtmeier_bohnenstengel_1981}, or $8.9\times10^9\,M_\odot$ from HIPASS \citep[][see corresponding global profile in Fig.~\ref{figure:globprof}]{koribalski_etal_2004}, both corrected for distance. This implies that we recover all of the \HI\ mass in the system with our KAT-7 observations. The \HI\ column density map is displayed in Figure~\ref{figure:coldens}. The column density contours are overlaid on the WISE multi-color image in Figure~\ref{figure:optoverlay}. Plainly, the outer \HI\ envelope is greatly extended beyond the bright optical disk. The size of the \HI\ disk is larger than seen previously in interferometric observations, as demonstrated in Fig.~\ref{figure:thingsoverlay} in which we compare with the high-resolution (but single-pointing) map from THINGS \citep{bigiel_etal_2010}. The improved recovery of outer disk \HI\ comes from a combination of factors: larger field of view per pointing, the mosaic observations that we have employed, and excellent sensitivity to low surface brightness emission \citep[see also][]{lucero_etal_2015}. It also leads to the KAT-7 observations having recovered substantially more \HI\ mass than the THINGS observations. In Fig.~\ref{figure:globprof} we have illustrated the reason for this in two ways: (i) the global \HI\ profile that would result by filtering the KAT-7 data cube with the VLA primary beam; (ii) the global \HI\ profile that results by masking the KAT-7 data cube with the THINGS moment-0 map. Clearly the field of view explains some of the \HI\ mass missed by THINGS, but not all. With KAT-7 we recover the same radial distribution of \HI\ as was implied by the single-dish Effelsberg map from \citet{huchtmeier_bohnenstengel_1981}, which we checked by smoothing the KAT-7 column density map to $9\arcmin$ resolution and performing a direct comparison.

In Figure~\ref{figure:optoverlay} it is apparent that the M~83 field contains some additional galaxies that are associated with M~83 itself. The companion galaxy visible to the east is NGC~5264. The companion to the north is cataloged as E444-78 (UGCA~365). There is also an object present in the outer \HI\ disk, cataloged by \citet{karachentseva_karachentsev_1998} as KK~208. We return to this interesting object in \S\,\ref{subsection:environment}.

We find an \HI\ mass for the northern galaxy (UGCA~365) of $M_\mathrm{HI}=2.7\times10^7\,M_\odot$ (assuming $D=5.26\,\mathrm{Mpc}$ based on 4 TRGB measurements in NED\footnote{The NASA/IPAC Extragalactic Database (NED) is operated by the Jet Propulsion Laboratory, California Institute of Technology, under contract with the National Aeronautics and Space Administration.}), and for the eastern galaxy (NGC~5264) we find $M_\mathrm{HI}=2.5\times10^7\,M_\odot$ (assuming $D=4.66\,\mathrm{Mpc}$ based on 3 TRGB measurements in NED).

\begin{figure}
\includegraphics[width=\hsize]{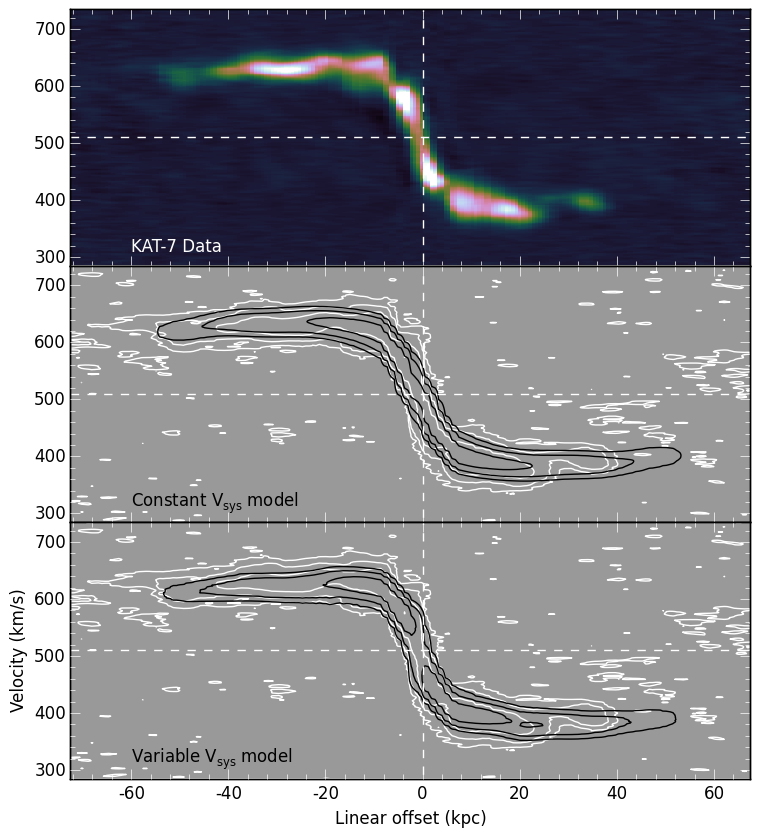}
\caption{Position-velocity (PV) diagram constructed by following the line of nodes defined by the PA of rings in the model shown in Figure~\ref{fig:vsysvar}. The top panel shows the data PV diagram, and the other diagrams show comparisons with the models presented in Figures~\ref{fig:vsysconst} and \ref{fig:vsysvar} respectively. White contours are from the \HI\ data cube, and black contours are from the model data cube. Contour levels start at $15\,\mathrm{mJy\,beam^{-1}}$ and increase by powers of four.}
\label{figure:pv}
\end{figure}

\begin{figure*}
\includegraphics[width=0.5\hsize]{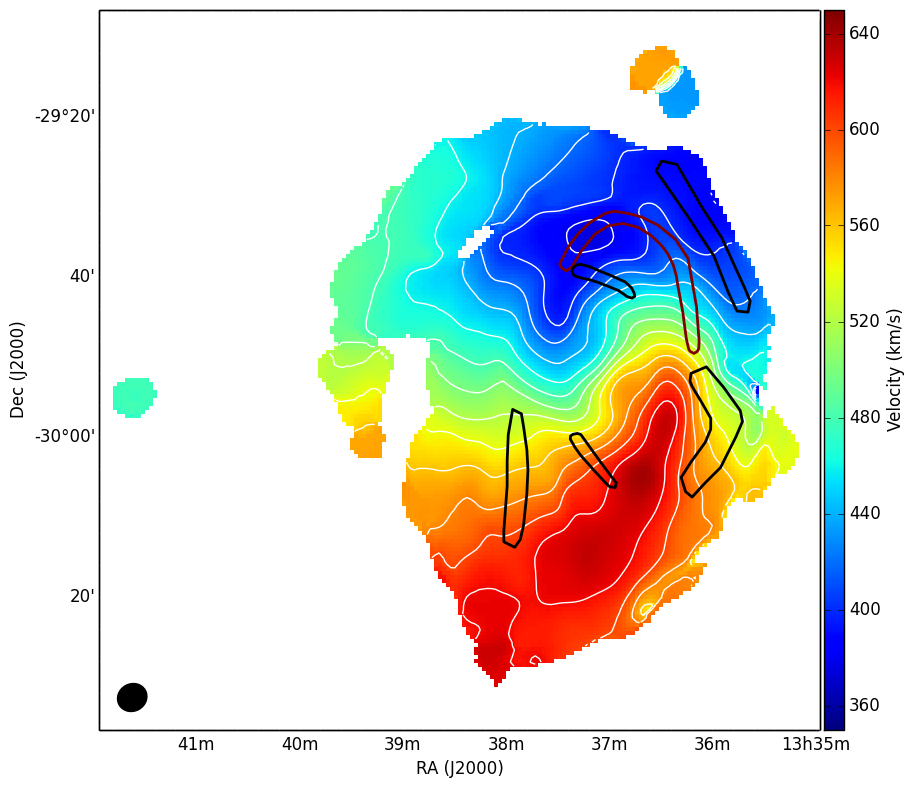}\includegraphics[width=0.5\hsize]{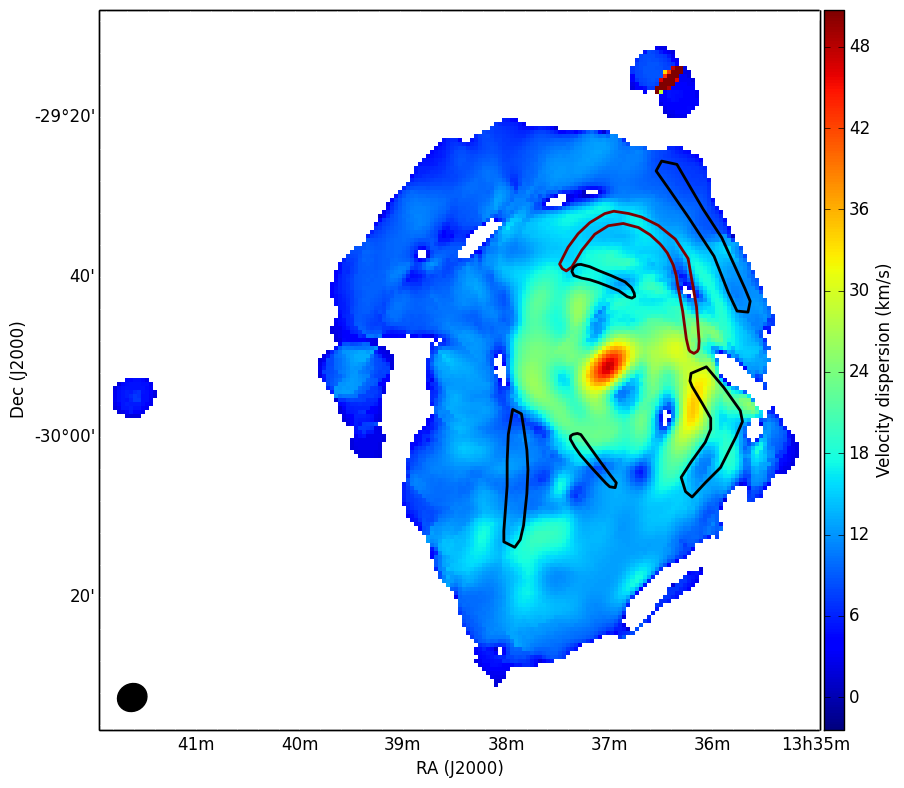}
\caption{Velocity field (left) and velocity dispersion (right), both clipped to only display values for pixels that have column densities above 5.6e18. The beamsize is shown in the bottom left of each panel with a black ellipse.}
\label{figure:velfielddisp}
\end{figure*}

The \HI\ velocity field and velocity dispersion of M~83 are displayed in Figure~\ref{figure:velfielddisp}. While the velocity field is very well behaved, it clearly shows signs of a substantial warp in the outer parts. Beyond the optical radius, the position angle twists to almost $90\degr$ from the orientation within the optical body. Twists in the isovelocity contours, particularly in the northern arm-like extension, betray the presence of substantial non-circular motions in the gas distribution. We model the structure of the \HI\ disk in \S\,\ref{subsection:trm}. The velocity dispersion map is discussed in some detail in \S\,\ref{section:outersf}, where we also consider star formation in the outer disk.

In the rest of this section we highlight some interesting aspects of the extended disk in M~83, and its connection to the larger environment. Since the \HI\ is very extended, we will address the morphology and kinematics far from the optical body. We will also discuss connections to the companion galaxies.

\subsection{Tilted ring modeling}\label{subsection:trm}

The moment maps of M83 show that the galaxy can be divided into a
fairly regular and symmetric inner part, corresponding with the
optical disk, and a more asymmetric and extended outer part. Of this
outer part the southern (receding) half seems to be more regular (both
in morphology and kinematics) than the northern (approaching) part.
The angular resolution of the data does not allow a detailed in-depth study
and mass-modeling of the rotation curve of M83. In this sub-section
we will concentrate on the dynamics of the outer disk and the presence
of non-circular motions. Uncertainties in the inner rotation curve are discussed
in \S\,\ref{section:rccompare}.

\subsubsection{The simplest tilted-ring model}

We use the intensity-weighted mean (IWM) first-moment map and the
tilted-rings method to explore the rotation curve of M83.  We use the
{\sc gipsy} task {\tt rotcur} and adopt rings with a width of
$100\arcsec$ (half the beam size), which Nyquist-samples the velocity
field.  We apply a cosine-weighting to the velocity field: each pixel
is given a weight equal to the cosine of the position angle with
respect to the major axis. We initially set the radial velocity
component to be zero.

We first constrain the position of the dynamical center and the value
of the systemic velocity. We run {\tt rotcur} with all parameters free
(except for the zero radial velocity) and find that within $R <
1000\arcsec$ the parameters are fairly stable.  For these radii, the
position of the centers of the rings scatters by around $30\arcsec$
(about one-sixth of a beam) around the optical center.  The systemic
velocity here equals $510\,\mathrm{km\,s^{-1}}$ with a scatter of
$4\,\mathrm{km\,s^{-1}}$.

At radii $r > 1000\arcsec$ the behaviour is different. Here the
position of the dynamical center changes with radius with a maximum
offset of $\sim 130\arcsec$ towards larger RA and $\sim 360\arcsec$
towards smaller declination. The scatter is however as large as the
offset, and for these outer rings the position of the dynamical center
cannot be determined independently. Here the systemic velocity values
decrease to $\sim504\,\mathrm{km\,s^{-1}}$ with a scatter of
$\sim5\,\mathrm{km\,s^{-1}}$.  These changes in center position and
systemic velocity indicate that the outer rings either have a
non-circular motion component or are not concentric with the inner
rings (or both).  We explore these possibilities later.

To get a first indication of the rotation curve we assume that the
center positions of the rings at small radii also apply to the outer
disk and we fix the position of the dynamical center to that of the
optical center ($13^h37^m00.9, -29\degr51\arcmin56\arcsec$).

A second run of {\tt rotcur} with the center position fixed, but the
remaining parameters free, shows a well-defined systemic velocity of
$510 \pm 2\,\mathrm{km\,s^{-1}}$ between 0 and $800\arcsec$.  Between
$800\arcsec$ and $1000\arcsec$ the velocity quickly drops by
$10\,\mathrm{km\,s^{-1}}$ and stays constant at a value of $500 \pm
1\,\mathrm{km\,s^{-1}}$ between $1000\arcsec$ and $2100\arcsec$.  We
again assume that the values found in the inner parts apply to the
entire disk.

\begin{figure*}
\includegraphics[width=\hsize]{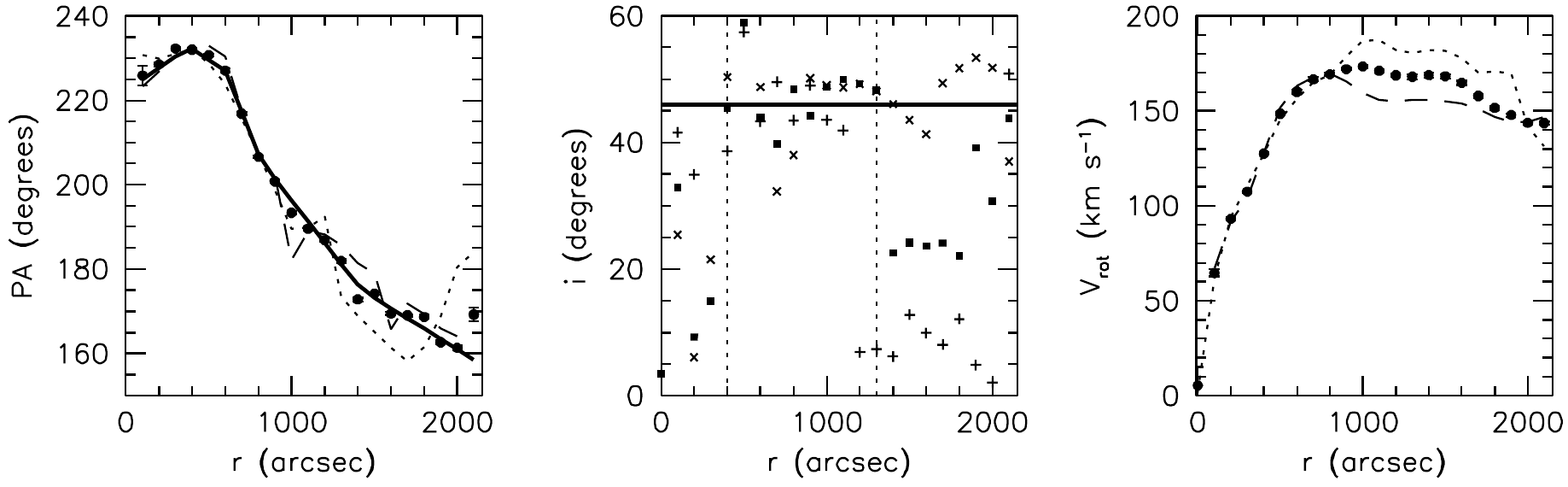}
\caption{Model parameters for the constant systemic velocity model. Left panel 
shows the distribution of PA. Filled circles are the PA values found
in a run with PA, inclination and rotation velocity as free parameters
using the entire disk. The dotted line shows the same for the
approaching side, the dashed line for the receding side. The thick
solid curve show the model PA that is used for subsequent runs. Center
panel: distribution of the inclination, where only inclination and
rotation velocity are free parameters and the PA is fixed to the
model shown in the left panel. Filled squares show the values derived
for the entire disk; $\times$-shaped crosses those for the approaching
side; $+$-shaped crosses those for the receding side. The two vertical
lines indicate the range over which the average value was determined,
as described in the text. Right panel: rotation curve, derived with
all other model parameters fixed. Filled circles show the rotation
curve for the full disk, the dotted and dashed lines those for the
approaching and receding sides respectively.}
\label{fig:vsysconst}
\end{figure*}

With rotation velocity, inclination and position angle as the only
remaining free parameters, we derive models for the entire velocity
field, as well as the approaching and receding sides.  For all three
models the position angle is very well determined, and largely
independent of which part of the velocity field is modeled. This is
shown in Fig.~\ref{fig:vsysconst}.  We describe the radial variation
of the position angle using a simple multi-line-segment model, and fix
this parameter in subsequent runs.

We re-run these three models, this time with only the rotation
velocity and the inclination as free parameters.  The model using the
entire velocity field does not show a clearly preferred value for the
inclination, with values ranging between $0\degr$ and almost
$50\degr$.  The approaching and receding sides models show a more
stable behaviour.  The approaching side shows two distinct inclination
ranges. At radii up to $1100\arcsec$ we find an average value of
$44 \pm 6$ degrees. At larger radii a steep drop in the inclination
results in values $8 \pm 3$ degrees. Such low values would lead to
deprojected rotation velocities between $\sim 500$ and $\sim
1000\,\mathrm{km\,s^{-1}}$ and it is clear these are not physical.
The receding side shows the opposite behaviour. Here, the inclination
yields unphysical values at radii $r < 900\arcsec$, while for larger
radii the inclination is more stable with a value $48 \pm 3$ degrees.

Comparing the results from the approaching and receding sides, we find
that the radial range between $400\arcsec$ and $1300\arcsec$ gives the
most representative values for the galaxy overall. Interior to
$400\arcsec$, the fit appears to be affected by beam smearing. Between
$r = 400\arcsec - 1300\arcsec$, we find an
average inclination value of $46 \pm 3$ (using the model based on the
entire velocity field and excluding the outlying value at $r =
500\arcsec$).  We adopt this value for the entire galaxy and find the
rotation curve as shown in Fig.~\ref{fig:vsysconst}.  The Figure also
shows the rotation curves of the approaching and receding sides using
the same assumptions. A position-velocity (PV) slice through a mock
observation of this model (formed using the \HI\ radial column density
distribution displayed in Fig.~\ref{fig:sd}) is compared with the data
in Fig.~\ref{figure:pv}.

For radii $r > 1000\arcsec$ we see a systematic and opposite deviation
of the rotation curves of the approaching and receding sides. In the
context of the purely rotational and azimuthally symmetric model,
such a deviation is an indication of an incorrectly chosen systemic
velocity or center position. In the case of M83, it more likely indicates that
there are asymmetries or non-circular motions in the disk that a
simple tilted-ring model cannot properly take into account.

\subsubsection{Models with non-circular motions}

\begin{figure*}
\includegraphics[width=\hsize]{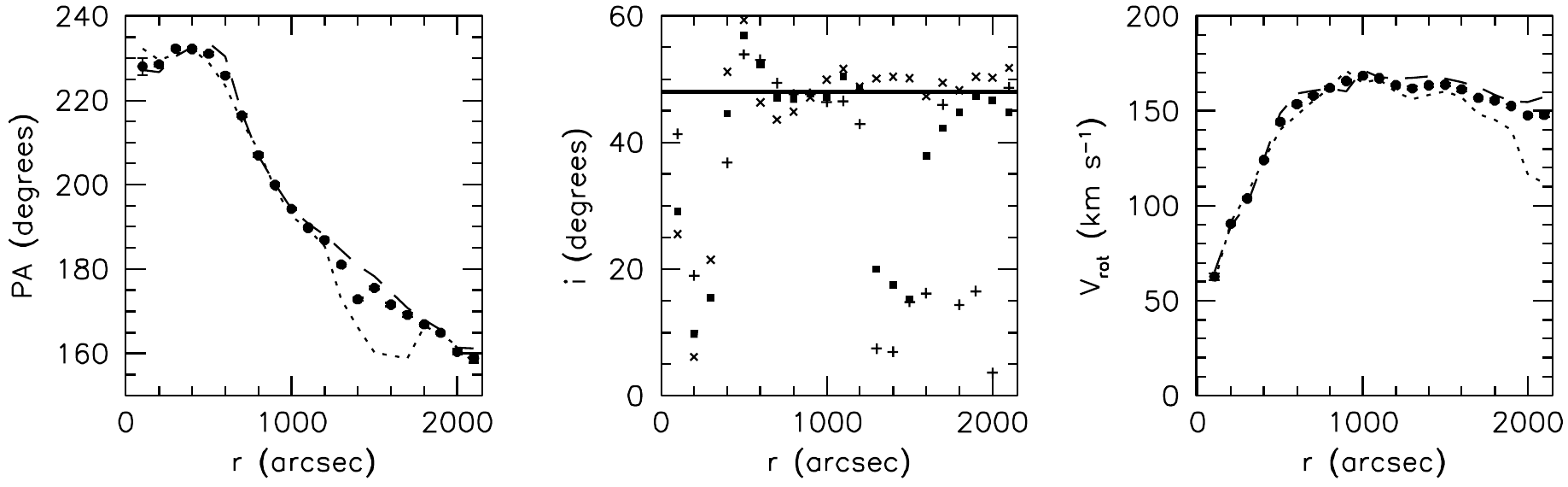}
\caption{Model parameters assuming a variable systemic velocity of $V_{\rm sys} = 510$ 
km s$^{-1}$ inside $r=900''$ and $V_{\rm sys} = 500$ km s$^{-1}$
outside $r=900''$. All symbols and lines are as in
Fig.\ \ref{fig:vsysconst}, except for the center panel, where the
thick horizontal line indicates the average value of the inclination
used in the modeling.}
\label{fig:vsysvar}
\end{figure*}

To explore the origins of these deviations we run a set of models
where we adopt $V_\mathrm{sys} = 510\,\mathrm{km\,s^{-1}}$ for $R \le
900\arcsec$ and $V_\mathrm{sys}=500\,\mathrm{km\,s^{-1}}$ for $R >
900\arcsec$. As above, we again find well-constrained position angle
values as a function of radius, and use a simple line-segment model to
describe these. The position angle values are shown in
Fig.~\ref{fig:vsysvar} and are fixed to the model values hereafter.

A new run with only the inclination and the rotation velocity as free
parameters again shows a large scatter in inclination. Ignoring the
obvious outliers and unphysical inclination values, we use the radii
between $600\arcsec$ and $1200\arcsec$ to determine an average value
of $48 \pm 1$ degrees.

The receding side inclination values suggest that this value is
representative for the galaxy. We therefore use this value to derive
the final rotation curves. Beyond a radius of $\sim 1000\arcsec$ the
approaching and receding side rotation curves still differ
substantially. Care should therefore be taken not to over-interpret
the rotation curve at these large radii. A PV slice through this
model is compared with the data in Fig.~\ref{figure:pv}. 

We investigate the uncertainty of the outer rotation curve by
quantifying the effect of center offsets and
non-circular motions in the outer disk. In particular, we explore what
non-circular motions or center offsets are needed to produce a model
that is consistent with a flat rotation curve with an amplitude of 165
km s$^{-1}$ beyond $r=1400''$. For a model where we allow the center
position to vary, we find that to produce a flat rotation curve in the
outer parts, the central positions of these outer rings only need to
change by $\sim 200''$, or less than one beam. Similarly, if radial
motions are used to make the rotation curves flat, only about
$10\,\mathrm{km\,s}^{-1}$
of radial motions are needed.

Comparing the various models derived in this Section, we therefore
conclude that under the various assumptions presented here the outer
rotation curve is uncertain by about $\pm 20$ km s$^{-1}$. These
unconstrained rotation velocities at large radii thus provide only
limited information about the dark matter distribution in the outer
disk of M83.

\subsection{Discussion of the RC and comparison with previous HI studies}\label{section:rccompare}

\begin{figure}
\centering
\includegraphics[width=90mm]{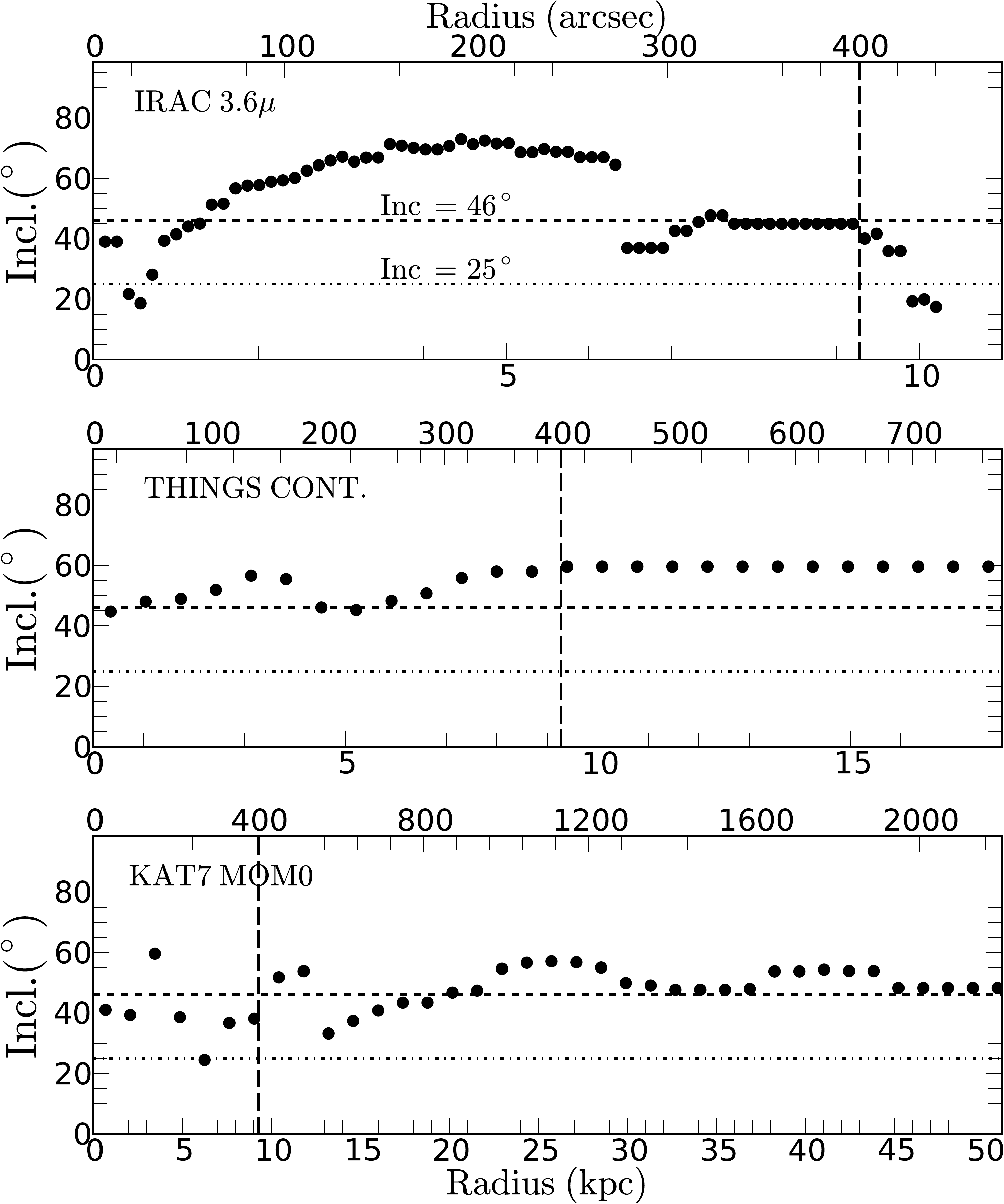}
\caption{Comparison of the inclinations derived from isophotal fitting of the IRAC 3.6 $\mu$m image, the THINGS (natural weighting) continuum image \citep{deblok_etal_2008} and the \HI\ total intensity map from our KAT-7 data. The horizontal dotted lines show inclinations of 25\degr\ and 46\degr\ and the vertical dashed lines show the radius of 400\arcsec.}
\label{fig:inc}
\end{figure}

The last section (see Figures~\ref{fig:vsysconst} and \ref{fig:vsysvar}) has shown clearly that while the variation of the PA from 230\degr\ to 160\degr\ is very well defined by the tilted ring analysis, the variation of the inclination for $r <$ 400\arcsec\ and $r >$ 1200\arcsec\ is much less constrained, being more or less constant $\sim$46\degr\ between those two radii. In order to try to constrain better the inclination, it was decided to use means other than the kinematics. Figure \ref{fig:inc} compares the inclinations derived  from isophote fitting of the IRAC 3.6 $\mu$m image \citep{barnes_etal_2014}, the THINGS (natural weighting) continuum image \citep{deblok_etal_2008} and the \HI\ total intensity map from our KAT-7 data. For the continuum, the THINGS data were adopted because of their better spatial resolution, which is important for the inner parts.

We first looked at the IRAC 3.6$\mu$m image. While the inclination is not well defined in the inner 50\arcsec\ (35.2\degr\ $\pm$ 9.8\degr), it tends toward more edge-on values for 50\arcsec\ $< r <$ 250\arcsec\ (65.0\degr\ $\pm$ 6.2\degr) and intermediate in the disk region for $r >$ 250\arcsec\ (43.3\degr\ $\pm$ 12.1\degr). However, this infrared image is probably not very useful to constrain the disk inclination since the isophotes between 50\arcsec\ $< \rm r <$ 250\arcsec\ trace the morphology of the inner bar rather than the disk, which explains the very high implied inclination values. For the continuum image, we obtain an inclination of 51.4\degr\ $\pm$ 5.0\degr\ for  $r <$ 400\arcsec\ and 55.5\degr\ $\pm$ 5.4\degr\ over the whole radius range. Finally, we consider a fit to the isophotes of our KAT-7 \HI\ total intensity map. It can be seen again that the inclination is not well constrained for $r <$ 400\arcsec\ (39.7\degr\ $\pm$ 10.4\degr) but is quite constant out to the largest radius (47.5\degr\ $\pm$ 7.6\degr).

From all of these comparisons, we conclude that a constant inclination of 24\degr\ as adopted by \citet{park_etal_2001} is not justified from our analysis. Moreover, such a face-on orientation would imply very large rotation velocities of $\sim$300 \kms\ at $r \sim$15\arcmin\, implying an absolute magnitude $M_B \sim -23$ \citep{persic_salucci_1991, persic_etal_1996}, which is very bright when compared to the absolute magnitude $\sim-21$, estimated for M83 at our adopted distance (see Table~\ref{table:m83properties}). The actual absolute magnitude is compatible with our assumed inclination of $46\degr$ and the corresponding maximum velocity $\sim$170 \kms.

\begin{figure}
\centering
\includegraphics[width=90mm]{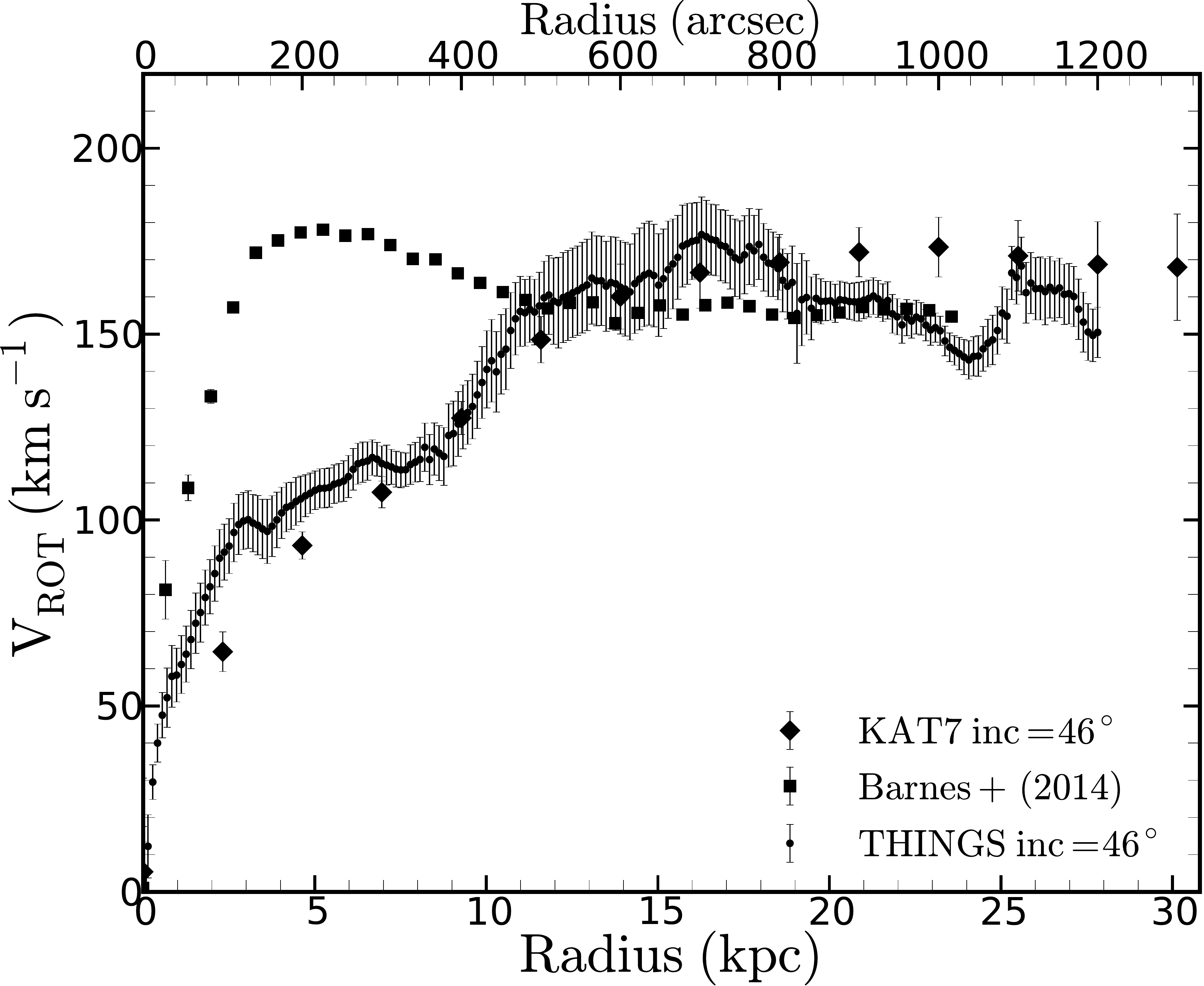}
\caption{Comparison of the KAT-7 rotation curve with the ones from THINGS \citep{deblok_etal_2008} and \citet{barnes_etal_2014}.}
\label{fig:comp}
\end{figure}

\begin{figure}
\centering
\includegraphics[width=90mm]{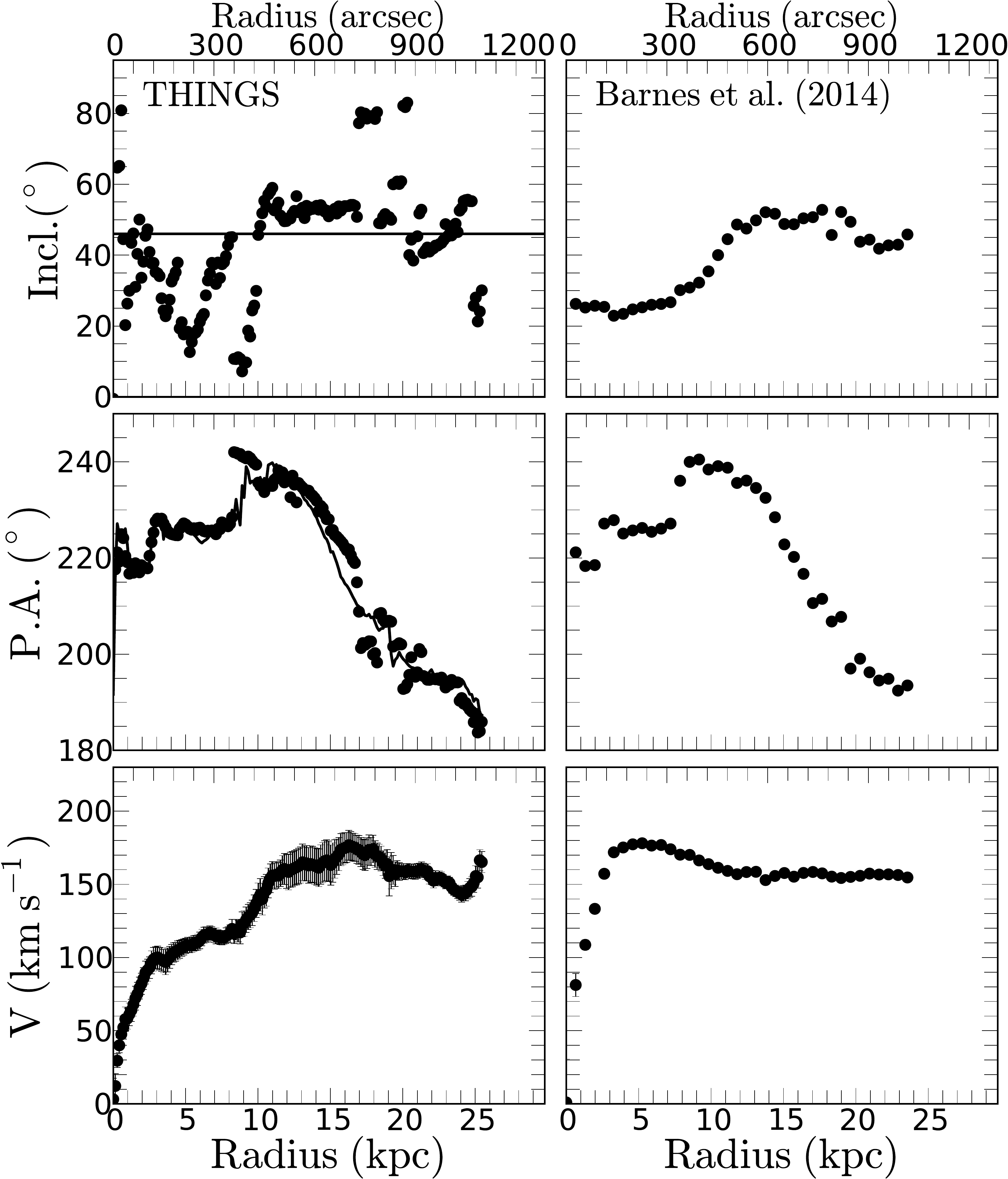}
\caption{Comparison of the kinematical parameters determined from the \HI\ data presented by \citet{deblok_etal_2008} and \citet{barnes_etal_2014}. On the left, the dots are from a {\sc rotcur} solution with $i$ and PA left free to vary. The continuous lines illustrate the adopted parameters. On the right, we show the $i$ and PA values that correspond to the rotation curve determined by \citet{barnes_etal_2014}.}
\label{fig:kin}
\end{figure}

In order to be able to consistently compare the available data, a constant inclination of 46\degr\ was adopted for the rotation curves derived from KAT-7 (this work) and one derived for the VLA (THINGS) data (note that \citet{deblok_etal_2008} did not derive a rotation curve for M83 in their paper). Figure \ref{fig:comp} compares those two rotation curves with the one derived by \citet{barnes_etal_2014}, also from VLA data. First, if we compare the KAT-7 and THINGS rotation curves, we see that they agree fairly well between 300\arcsec\ and 800\arcsec. Some beam smearing is apparent for $r <$ 300\arcsec, and the KAT-7 rotation velocities are $10-20$ \kms\ higher around 1000\arcsec\ before becoming consistent with the outermost THINGS value around 1100\arcsec. On the other hand, while all three rotation curves are in agreement for $r >$ 400\arcsec, the velocities from\citet{barnes_etal_2014} are much higher in the inner parts. This can be understood easily through comparison with Figure~\ref{fig:kin}. While the rotation curves from both KAT-7 and THINGS were developed using a fixed inclination of 46\degr, \citet{barnes_etal_2014} allowed the inclination to go down to $\sim$20\degr\ in the inner parts, resulting in much higher rotational velocities.

Because of the uncertainties in the inclination for $r <$ 400\arcsec\ and $r >$ 1200\arcsec, no attempt will be made in this paper to derive mass models for M83 since the inner parts (strong gradient) are crucial to distinguish between cored (e.g. ISO) and cuspy (e.g. NFW) dark matter profiles and the outer parts for MOND models.

\subsection{Environment and interactions}\label{subsection:environment}

By comparing the data velocity field with the velocity field derived from our tilted ring model, we can identify deviations from pure circular rotation and the simple geometry incorporated in the modeling. In the central parts of the galaxy, the velocity deviations are small ($\approx0\pm8\,\mathrm{km\,s^{-1}}$) and consistent with the star forming disk being well behaved. At intermediate radii, there is a spiral shape to the velocity residuals that is consistent with the outer \HI\ arm morphology. These arms are oriented with a pitch angle opposite in sign with respect to the optical arms (i.e., counter-clockwise winding, as opposed to clockwise winding in the inner disk and far outer parts). This may be the sign of a past retrograde interaction. In the residual velocities we also note a large-scale gradient (magnitude $\pm50\,\mathrm{km\,s^{-1}}$) from east to west in the far outer parts. This may be a consequence of the apparent offset in rotational velocity from approaching to receding side (see Fig.~\ref{fig:vsysconst}), and/or may indicate that a small shift in the dynamical center is appropriate for the outer rings (see \S\,\ref{subsection:trm}).

The stellar stream that has previously been cataloged as KK~208 is visible in Figure~\ref{figure:coldens}. The stream's shape and extent is outlined for clarity in the optical image, as well as in the \HI\ column density map, velocity field, and velocity dispersion map. The stream clearly has a recognizable signature in the kinematics of the \HI\ disk. Systematic twists in the isovelocity contours coincide with the location of the stream, indicating that the stream has had a gravitational influence on the outer disk gas. This provides additional evidence that the stellar stream is the remnant of the object that interacted with the outer disk of M~83 \citep[see also][]{barnes_etal_2014}, causing its disturbed outer disk morphology and kinematics, and potentially creating the conditions needed to initiate the prominent outer-disk star formation. With future deep \HI\ followup observations (e.g. with MeerKAT), the details of the \HI\ in this region can be studied at similar surface brightness sensitivity but much higher angular resolution, shedding more light on the environmental effects that have helped to shape M~83.

\subsection{Outer disk star formation and velocity dispersion}\label{section:outersf}

\begin{figure*}
\includegraphics[width=\hsize]{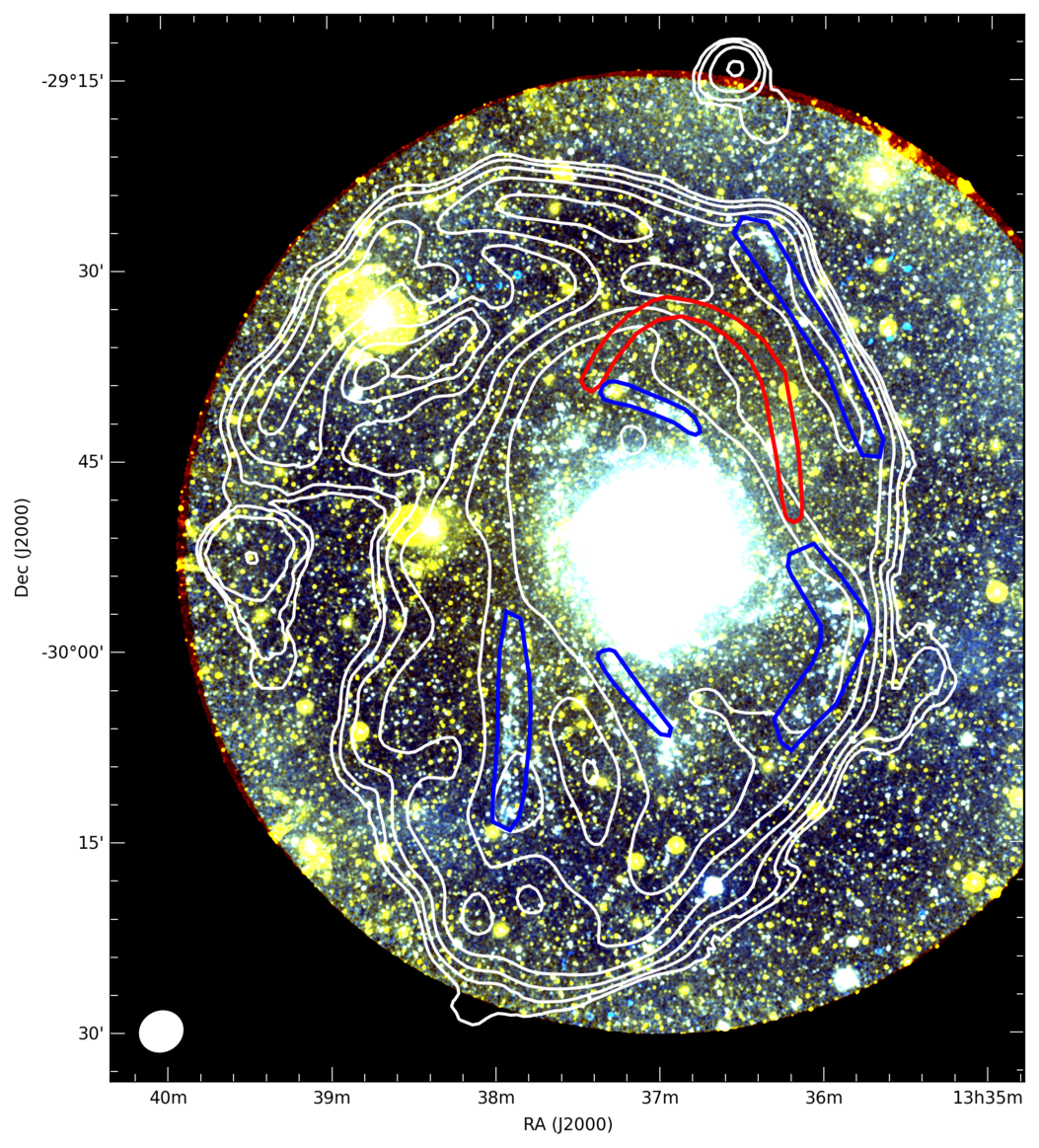}
\caption{Overlay of KAT-7 \HI\ column density distribution on a false-color image formed from GALEX NUV and FUV images, both smoothed to $10\arcsec$. The colors are chosen such that FUV emission appears blue. The \HI\ column density contour levels are the same as in Fig.~\ref{figure:coldens}. The beam size is shown with a white ellipse in the bottom-left corner.}
\label{fig:galex}
\end{figure*}

With our \HI\ data at its relatively coarse angular resolution, we are only able to broadly compare the distribution of outer-disk star formation to the neutral gas distribution. As compared with the strong spatial correlation described by \citet{bigiel_etal_2010}, we can only confirm that the regions of star formation are co-located with local enhancements in \HI\ column density. An overlay of the neutral gas distribution on deep GALEX imaging is shown in Fig.~\ref{fig:galex}. Note that the two stellar stream segments most closely connected to the main optical disk are seen in GALEX as outer disk spiral arms, as well as in the deep optical image (Fig.~\ref{figure:coldens}). The loop-like stellar stream is not visible with GALEX. This is consistent with the loop structure having a different origin than the outer-disk star forming arms.

On the other hand, a large mass of \HI\ is present in the outer parts of M~83 and {\it unassociated} with ongoing star formation. For the most part, this gas distribution was not recovered with the much higher angular resolution VLA observations (see Fig.~\ref{figure:thingsoverlay}). Much of the discrepancy in recovered \HI\ mass is attributable to a lack of surface brightness sensitivity in the THINGS data (Fig.~\ref{figure:globprof} and \S\,\ref{subsection:properties}). We therefore conclude that the bulk of the outer disk neutral gas is diffuse and not clumpy on scales recovered by the VLA, supplemented with some \HI\ overdensities tracing arm structures that became gravitationally unstable, allowing outer disk star formation to proceed.

We conclude this discussion with a brief look at the \HI\ velocity dispersion (Fig.~\ref{figure:velfielddisp}). Enhancements in velocity dispersion may be related to star formation properties and their impact on the ISM \citep[e.g.,][]{ianjamasimanana_etal_2015}, providing additional leverage on the relationship between gas and star formation. However, we find that in our low-resolution KAT-7 \HI\ data, systematic resolution effects provide a dominant contribution to the velocity dispersion in particular regions. At the innermost radii, beam smearing causes the highest dispersion values ($\gtrsim40\,\mathrm{km\,s}^{-1}$; see Fig.~\ref{figure:velfielddisp}). At intermediate radii, many of the regions of high velocity dispersion tend to align with locations where tilted rings in our unconstrained models (\S\,\ref{subsection:trm}) overlap due to projection effects. This is borne out by the appearance of the line profiles in those locations, which are reminiscent of the superposition of more than one velocity profile within the large beam rather than broadening of a single profile. An analysis of the \HI\ velocity dispersion must therefore await higher angular resolution observations. Although M~83 is not part of the MHONGOOSE sample (because mosaicing is required), we emphasise that our KAT-7 data provide a strong case for a MeerKAT early science program focused on deep high-resolution mapping of the \HI\ in this system.

\subsection{Disk edge}

A prominent feature of the \HI\ disk of M~83 is the sharp edge to the gas distribution, which is clearly visible as a close packing of the outermost contours in e.g. Fig.~\ref{figure:optoverlay}. This edge may be an indication of dynamical effects in the environment, ionization effects from the IGM, or some other physical effect.

Before discussing possible physical origins for this morphological feature, we first investigate whether the apparent sharp edge in the \HI\ disk could be caused by technical issues. First we test whether the mask applied in the creation of the column density image could artificially truncate the edge of the \HI\ disk. We have repeated the same procedure for creating the column density map, but using a mask expanded in all three coordinates of the input data cube. Specifically, we took the binary 3D mask and smoothed it (using a 7-pixel boxcar kernel in velocity, and a Gaussian beam with FWHM=$450\arcsec$ -- twice the size of the synthesized beam -- in the angular dimensions). We then formed a new binary 3D mask by clipping the smoothed result at the 50\%-intensity level. Visual inspection confirmed that the new mask provides much more empty space at the outer edge of the disk. Recreating the column density map, using this new mask in the summation step, led to very little difference in the outer disk morphology. We conclude that the masking step has not artificially sharpened the disk edge.

As a second test, we have formed a total intensity image using a completely different approach that does not impose any morphological constraints. In this procedure, we identify the peak velocity in each spatial pixel, and shift the corresponding spectrum so that their peaks are all aligned at the same velocity ($0\,\mathrm{km\,s}^{-1}$). In this resulting ``shuffled'' cube, we integrated the central 45 velocity pixels ($67.5\,\mathrm{km\,s^{-1}}$) without any masking. This leads to an integrated intensity map with improper values in the central parts (where typical velocity widths exceed the summation range used here), but with accurate morphology in the outer parts, including a sharp edge similar in behavior to what is observed in our adopted column density map (and particularly on the western edge of the disk). This confirms that the sharpness of the \HI\ disk is robust to the details of creating the column density map.

A final technical point to consider is the possibility of missing flux on large angular scales due to a lack of short spacings. As highlighted by \citet{lucero_etal_2015}, the sensitivity of KAT-7 to emission on large scales is particularly good for an interferometer. Moreover, we have demonstrated in \S\,\ref{subsection:properties} that the total \HI\ mass that we have determined is fully recovered with respect to single-dish measurements. Yet locally, some missing flux is possible. Based on inspection of the column density map constructed using the expanded mask described above, we can identify regions of negative flux adjacent to the disk edge in some regions, and the histogram of column density values is non-Gaussian at negative values. The minimum column density value is $-1.96\times10^{19}\,\mathrm{cm}^{-2}$, or only $-3.5\sigma$ with respect to the nominal column density sensitivity in Table~\ref{table:cubeprops}. The integral of negative column densities only comes to $0.3\%$ of the integral of positive column densities. We consider it unlikely that this slight negative bowl can be responsible for the sharp edge, but cannot exclude that it has an impact on the morphology in localized regions. We discuss the possible effect as we proceed.

\begin{figure*}
\includegraphics[width=\hsize]{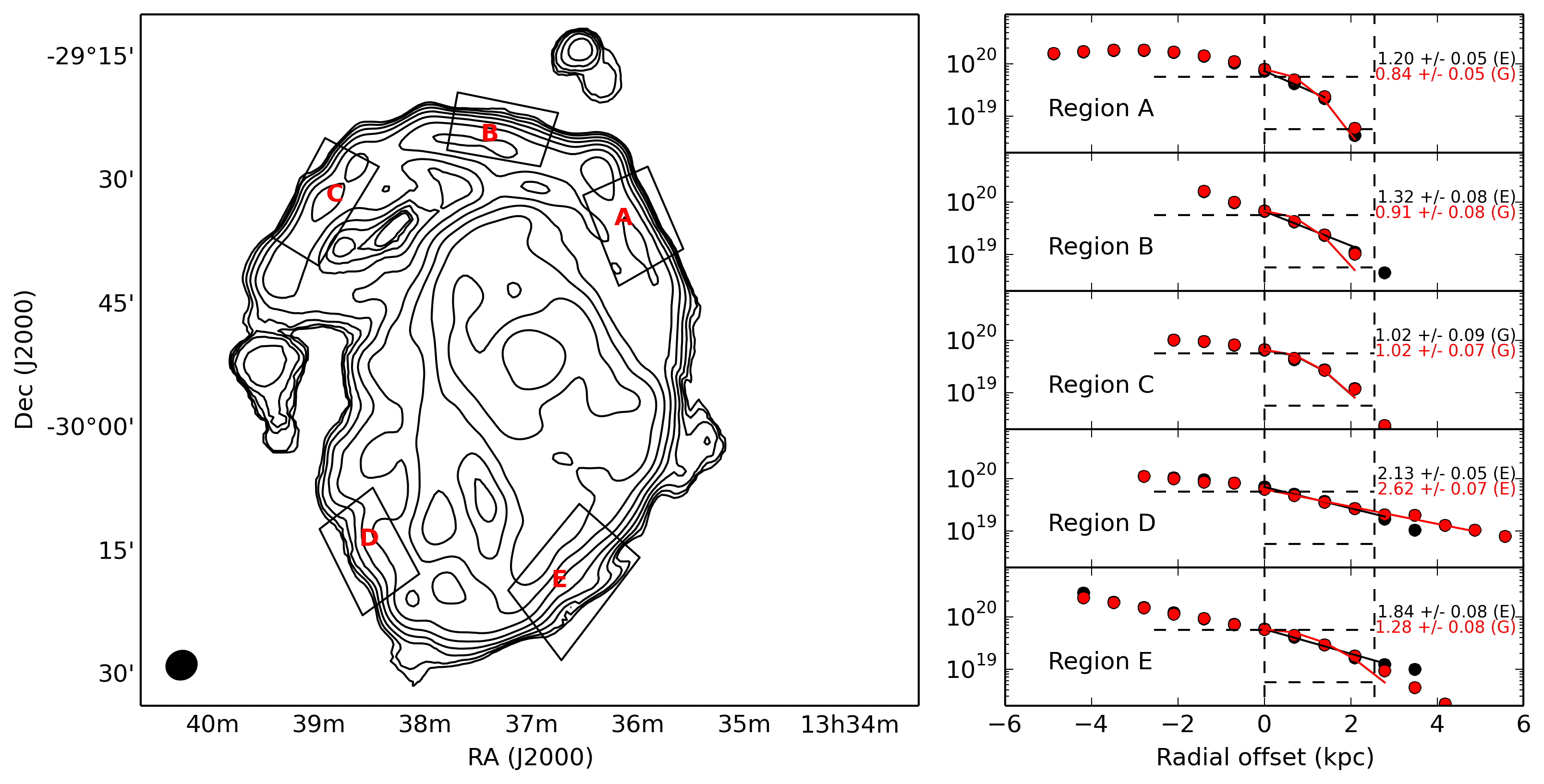}
\caption{Sharpness of \HI\ disk edges. The left panel shows the column density map, with 5 regions indicated. The average run of column density perpendicular to the disk edge for each region is shown in the right panels. The points are from the adopted column density map ({\it black}), and from the one created using an expanded mask ({\it red}). Lines show fits to the data points with characteristic length scales as indicated in the legend of each panel. Here, ``G'' indicates a Gaussian fit and the given value is the fitted value of $\sigma$. An ``E'' indicates an exponential fit and the given value is a scale length. In both cases the values are given in kpc. The typical length scale corresponding to the synthesized beam size is shown by the vertical dashed lines. The zeropoint of the radial offset is at the point where the average column density crosses the $N_\mathrm{HI}=5.6\times10^{19}\,\mathrm{cm}^{-2}$ level ({\it upper horizontal dashed line}), and the lower horizontal dashed line indicates the $N_\mathrm{HI}=5.6\times10^{18}\,\mathrm{cm}^{-2}$ level. In most cases the average column density drops by an order of magnitude within about a beamwidth, or $2-3\,\mathrm{kpc}$. The beamsize itself is shown with a black ellipse in the bottom-left corner of the lefthand panel.}
\label{figure:edgefig}
\end{figure*}

To quantify the structure at the edge of the disk, we have isolated five regions where the local morphology is relatively regular. Due to the large-scale and prominent asymmetries in the outer disk, we cannot simply study an azimuthally averaged radial profile. Instead, for each region we have constructed an average profile perpendicular to the disk edge (identified by tracing the $N_\mathrm{HI}=10^{19}\,\mathrm{cm}^{-2}$ contour). These average profiles are shown in Figure~\ref{figure:edgefig}. Each average profile was fitted with both an exponential and a Gaussian function (between the $N_\mathrm{HI}=5.6\times10^{18}\,\mathrm{cm}^{-2}$ and $N_\mathrm{HI}=5.6\times10^{19}\,\mathrm{cm}^{-2}$ levels) to obtain characteristic length scales. In each case, whichever functional form led to a fit with a lower covariance was adopted. The fitted functions are shown with the data points in Fig.~\ref{figure:edgefig}. It is remarkable that in all regions the characteristic length scale is very short: in most regions $\lesssim2\,\mathrm{kpc}$ (cf. the beamsize of our observations, $\approx5\,\mathrm{kpc}$). In fact the column density profiles demonstrate that the average column density drops by an order of magnitude within such a characteristic length scale.

This kind of sharp disk edge is very similar to the situation in NGC~3198 \citep[see][]{maloney_1993}, from deep VLA observations with similar column density sensitivity compared to our KAT-7 data. In NGC~3198, too, the azimuthally-averaged \HI\ column density distribution drops sharply ($\sim$order of magnitude within 3~kpc) at the edge of the disk, from $N_\mathrm{HI}=5\times10^{19}\,\mathrm{cm}^{-2}$ to $4\times10^{18}\,\mathrm{cm}^{-2}$. This implies a disk sharpness of $\Delta R/R\lesssim0.1$ where $\Delta R$ is the characteristic length scale of the disk edge. In the case of M~83, we have $\Delta R/R\lesssim0.03$. \citet{maloney_1993} models the \HI\ edge in NGC~3198 as an ionization effect; the intergalactic radiation field photoionizes the rarified gas in the far outer parts up to a characteristic column density consistent with that at the edge of NGC~3198, and, we now see, of M~83 as well. An implication of the photoionization interpretation is that the outer disk gas distribution is smooth, not clumped into clouds. This is consistent with our picture of the outermost \HI\ in M~83 that we developed when contrasting the VLA THINGS and KAT-7 datasets.

An alternative mechanism for producing a sharp disk edge is a ram pressure interaction with the surrounding IGM, as has been proposed for NGC~300 by \citet[][see also references therein]{westmeier_etal_2011}. If M~83 is moving through the IGM at a close to face-on trajectory, the sharp disk edge that we observe could represent the truncation radius where ram pressure dominates over the stabilizing gravitational pressure. Beyond that radius, no equilibrium is possible in this picture regardless of the vertical gas distribution. However, just inside the truncation radius the transition between dominance of ram pressure and the gravitational restoring force would be a function of the vertical gas distribution and this would then be responsible for the detailed morphology of the steep decline in \HI\ column density at the disk edge. Because we are unable to develop a reliable model of the dark matter halo in which M~83 resides (see \S\,\ref{section:rccompare}), we cannot constrain a quantified estimate of the impact of ram pressure in a similar fashion to \citet{westmeier_etal_2011}. However, we do note that for the conditions simulated in that paper (IGM density, and relative velocity of galaxy and IGM), the predicted transition radius occurs where the \HI\ column density declines to $N_{\mathrm{HI}}\sim10^{19}\,\mathrm{cm}^{-2}$, which is also the characteristic value at the \HI\ disk edge in M~83.

Finally, we point out that the sharpness of the \HI\ disk seems slightly enhanced toward the northern side of the galaxy, as compared to the south side of the galaxy. The sharper cutoff is visible by eye when inspecting the column density contours, and is also reflected in the shorter characteristic scale lengths presented in Fig.~\ref{figure:edgefig} for regions A, B, and C when compared to those in regions D and E. In the ionization scenario, this may be a result of a higher extragalactic ionizing photon flux from the the northern direction.
Alternatively, if ram pressure is the dominant effect in the outskirts of M~83, a northern component to the 3D trajectory of the galaxy might be indicated rather than a purely face-on interaction with the IGM.
Sensitive observations of the extended \HI\ distribution in nearby galaxies will be one of the outcomes of the MHONGOOSE survey, and it may become possible to draw a correspondence between disk edge sharpness and the properties of the environment, including the ionizing photon flux encountered by the sample galaxies from various directions.

\section{Radio continuum polarization and magnetic fields}\label{section:polarization}

Our initial observation of M83 was a broadband full-polarization continuum track, which we followed up with the line observations described in \S\,\ref{section:analysis}. In future, the MeerKAT correlator will be capable of creating both the broadband dataset simultaneously with the high velocity resolution line band. This will allow the study of the \HI\ structure and kinematics, while also probing the magnetic fields within the same objects.

To characterize the ordered component of the magnetic field in M83, we have created a series of narrowband images in Stokes $Q$ and $U$ as described in \S\,\ref{section:wideband}, and processed them using the Rotation Measure (RM) Synthesis \citep{brentjens_debruyn_2005} technique. This procedure performs a Fourier transform operation to translate $Q,U(\lambda^2)$ into $Q,U(\phi)$. Here $\lambda$ is the observing wavelength and $\phi$ is the so-called Faraday depth, which corresponds to the Faraday rotation measure in simple situations.

\begin{figure}
\includegraphics[width=\hsize]{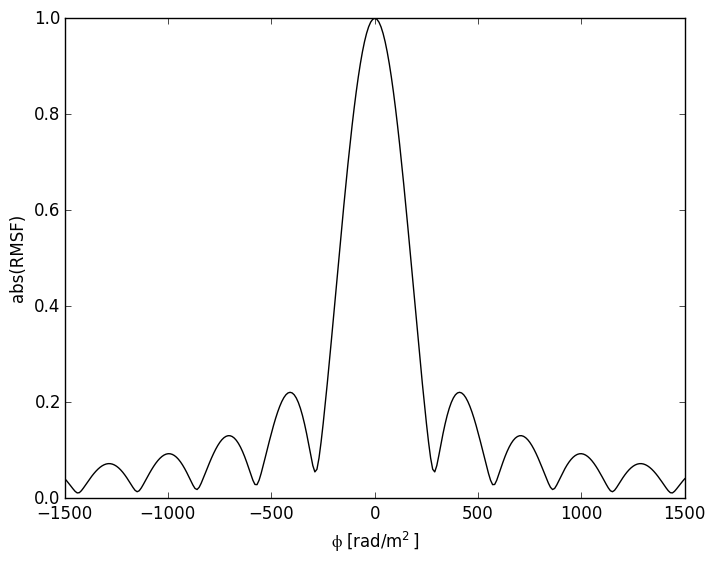}
\caption{The RMSF resulting from the KAT-7 continuum data presented in this paper.}
\label{fig:rmsf}
\end{figure}

The instrumental response function describing the ability to recover polarized emission at a precise Faraday depth is commonly referred to as the Rotation Measure Spread Function (RMSF). The RMSF from this wideband observation with contiguous frequency coverage is well behaved, as shown in Figure~\ref{fig:rmsf}. The FWHM of the main peak is $320\,\mathrm{rad\,m}^{-2}$, and the sidelobes are fairly low in magnitude ($\lesssim20\%$). Due to the relatively low frequencies that we have employed, the maximum scale of RM structure that can be recovered is only $73\,\mathrm{rad\,m}^{-2}$, less than the FWHM of the RMSF. This means that we are insensitive to Faraday thick emission (i.e. polarized emission that is present from a broad range of Faraday depths), which would originate for example from a medium extended along the line of sight that is both emitting polarized synchrotron radiation and causing Faraday rotation \citep[see, e.g.,][]{sokoloff_etal_1998}. In effect, our wideband KAT-7 observations are only capable of recovering Faraday screens. For comparison with the polarization properties at higher radio frequencies, we refer the reader to \citet{frick_etal_2015}.

\begin{figure*}
\includegraphics[height=0.45\hsize]{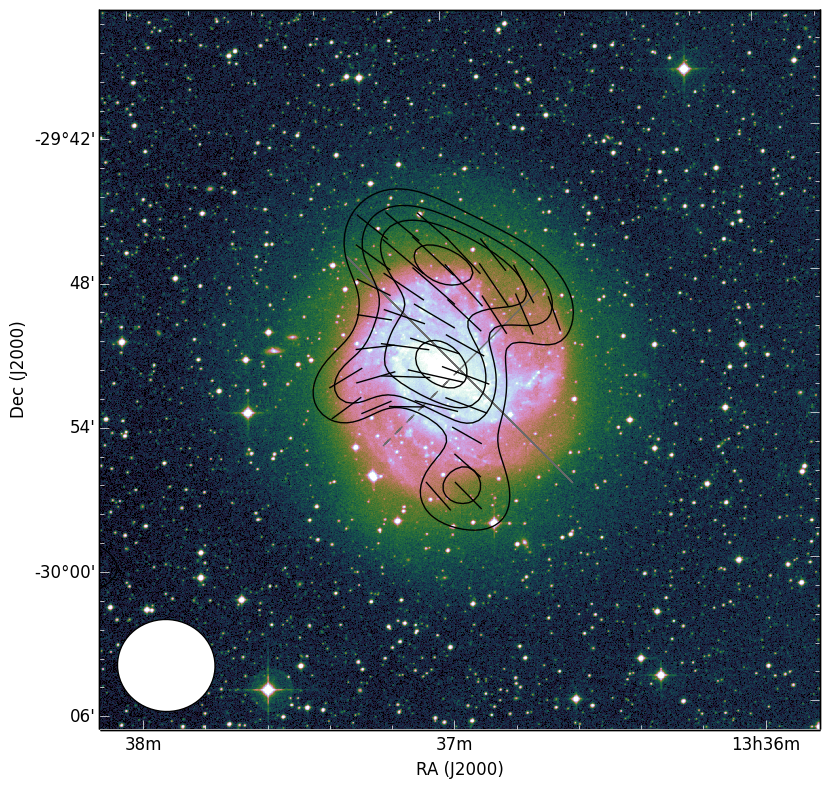}\includegraphics[height=0.45\hsize]{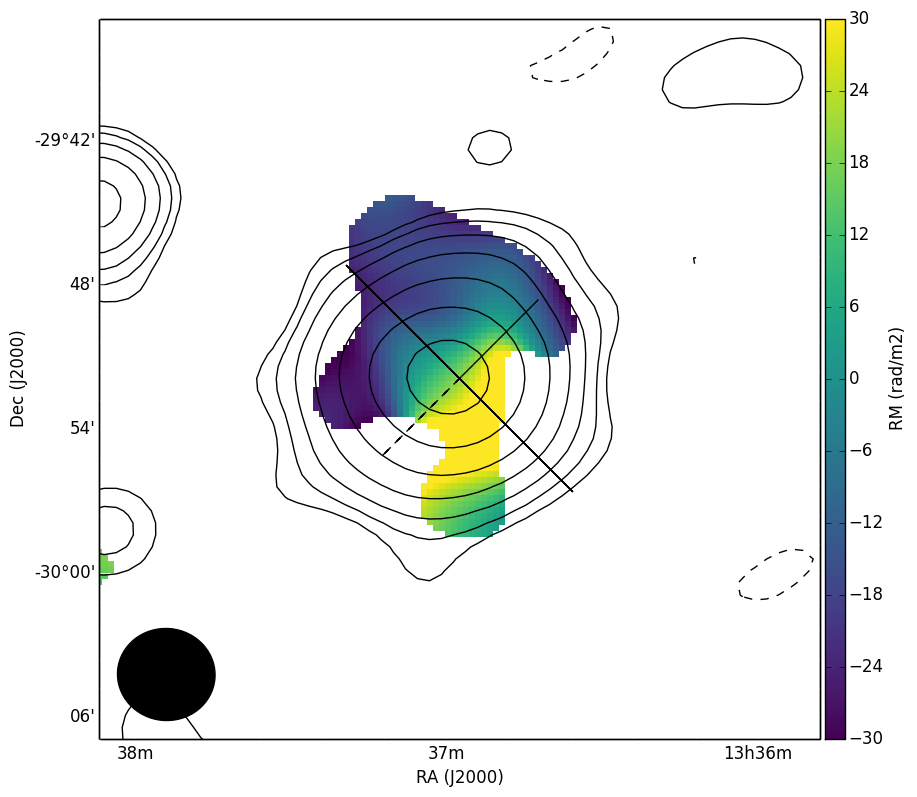}
\caption{Polarization map and magnetic field vectors (left) and associated Faraday RM (right). In the left panel the contours are at 8,10,12,15 mJy\,beam$^{-1}$ and the background is the POSS2/UKSTU Red image. The dark gray lines indicate the major (solid) and minor (dashed) axes; the kinematically receding side is to the southwest. The magnetic field vectors have been corrected for Faraday rotation, using the values in the right-hand panel. In the right panel the contours correspond to the total intensity continuum radiation, and are the same as in Figure~\ref{figure:continuum}.}
\label{fig:polarization}
\end{figure*}

After RM Synthesis, a cube is obtained that has Faraday depth as its third axis. Each frame contains the amount of polarized emission that has experienced a certain amount of Faraday rotation. This cube can be interpreted similarly to an \HI\ cube. In particular, we extract the peak value along each spectrum to obtain the linear polarization map, and record the corresponding Faraday depth at the location of the peak polarization to generate a RM map. This ``moment'' analysis was justified by inspecting the Faraday spectra to ensure that there were no complicated features (emission present over a broad range of Faraday depth). This simple approach therefore provides a reasonable reproduction of the polarized emission at these frequencies. The corresponding results are shown in Figure~\ref{fig:polarization}.

The polarized emission recovered in this way from our KAT-7 data is found to be distributed in an asymmetric fashion across the optical disk, in the sense that the approaching side (see Fig.~\ref{figure:velfielddisp}) is relatively highly polarized, but the receding side shows very little polarization. This large-scale pattern is in excellent agreement with many other spiral galaxies observed in polarization at similar frequency \citep[see][]{heald_etal_2009,braun_etal_2010,vollmer_etal_2013} and is consistent with early polarization mapping of M~83 at similar radio frequency \citep{sukumar_allen_1989}; see also \citet{neininger_etal_1993}. The orientation of the magnetic field vectors from the KAT-7 data is largely as expected, despite the low angular resolution, in that it follows the optical spiral pattern to a large degree. By comparison to the magnetic field distribution presented by \citet{frick_etal_2015}, we can see that the northern region of M~83 is dominated by magnetic fields tracing the spiral pattern, while the eastern side is dominated in polarization by magnetic fields following the bar orientation. At the easternmost edge, our derived magnetic field orientation appears at first glance to be contradictory to the polarization vectors presented by \citet{sukumar_allen_1989}, but on further inspection the apparent difference seems to be resolved when taking into account the rotation measure values in that region. To the south, the little polarization that is present at these low frequencies again follows the spiral pattern.

Note that the resulting average magnetic field orientation ($209.8\degr\pm3.7\degr$) is approximately parallel to the major axis (PA$\approx225\degr$; see Fig.~\ref{fig:vsysconst}). This seems to be consistent with the prediction by \citet{stil_etal_2009}, who demonstrate that an unresolved and idealized galaxy should show an average polarization angle (perpendicular to the mean magnetic field direction) oriented along the minor axis. The agreement between prediction and reality in this case is somewhat surprising given the substantial asymmetries in the polarized intensity from M~83 that are caused by the strong bar pattern, and strong localized depolarization that is systematically associated with the receding major axis as seems to be the case throughout the spiral galaxy population \citep[e.g.,][]{braun_etal_2010}. Whether this consistency with predictions is broadly true at these frequencies in a larger sample of galaxies remains to be seen, but will be important for interpretation of weak gravitational lensing signals that depends on polarization for a reference orientation \citep{brown_battye_2011}. Polarization observations at higher frequencies are important for unresolved galaxies, to avoid systematic effects such as depolarization biasing the recovery of the average polarization angle.

The rotation measure map corresponding to the polarized intensity is also presented in Fig.~\ref{fig:polarization}. We note that the foreground Milky Way rotation measure value is $-28.5\pm10.7\,\mathrm{rad\,m^{-2}}$ \citep{oppermann_etal_2015}. The bright approaching side shows Faraday rotation at a value similar to the foreground value, but the faint receding side displays an overall positive Faraday depth contribution. While our new rotation measure map is broadly consistent with the higher-frequency rotation measure maps from previous efforts, particularly given its lower angular resolution, there are differences that are possibly related to our lower observing frequency that probes the magnetic field at larger vertical distance from the midplane where depolarization effects are weaker. Taken together with the higher frequency rotation measures from \citet{neininger_etal_1993} and \citet{frick_etal_2015}, the change in sign of RM across the minor axis could suggest a large-scale field with a change in magnetic field directionality above the disk. Since the RM errors are dominated by low signal-to-noise (particularly in the receding side of the galaxy), deeper observations will be required to lower the uncertainties and confirm that the Faraday rotation measures indicate a large-scale ordered field in the disk-halo region.

We conclude by noting that future polarization observations of distant galaxies will have similar physical resolution as our current KAT-7 observation of M~83. The angular resolution here is $244\arcsec\times230\arcsec$. Thus we can expect that anticipated observations with greatly improved resolution ($\approx5\arcsec$) will recover the magnetic field structure on the same physical scales, but at redshift $z\approx0.06$. Through the detailed study of nearby objects like M~83 and comparison with sensitive but low angular resolution data from KAT-7, we are better able to understand future observations of distant targets.

\section{Conclusions}\label{section:conclusions}

We have presented new observations of the nearby grand design spiral galaxy M~83, performed with the KAT-7 radio interferometer in South Africa. Located on the site of the forthcoming MeerKAT radio telescope, and later SKA1-MID, KAT-7 is not only serving as an engineering testbed for the SKA-Mid precursor, but is also helping to develop the science case for MeerKAT survey projects such as MHONGOOSE.

With our new observations we have found that the \HI\ distribution of M~83 is more extended and diffuse than previously recognized. While outer disk star formation is closely coupled with clumpy, small-scale \HI\ features, a more extended and diffuse neutral gas component is also present. The morphology and kinematics of the extended \HI\ are consistent with a tidal interaction in the past, and we suggest that the northern stellar stream (previously cataloged as KK~208) may be the remnant of the interloper responsible for the disturbance. We also study the sharp edge of the \HI\ distribution and conclude that its abrupt termination (column density declining by an order of magnitude within a couple of kpc) is not an artifact but reflects a real change in the state of the ISM material in the far outer parts. Whether variations in the sharpness at various locations around the disk reflect directional variations in the ambient intergalactic ionizing radiation field or instead reflect the direction of motion of M~83 through the IGM remains to be tested with future observations.

The magnetic field distribution in M~83 was also probed with our KAT-7 observations. We find that the morphology of the polarized radiation can be understood in the context of both higher-resolution and higher-frequency polarization observations, and in comparison with the polarization properties of the broader spiral galaxy population. It may not always be the case that unresolved (or nearly unresolved) galaxies have a close relationship between optical position angle and the position angle of the linear polarization vector; asymmetries within the galaxy such as bars and systematically depolarized regions may serve to change the averaged polarization angle.

These observations help to set the stage for the MHONGOOSE survey, which will perform similarly sensitive \HI\ observations at much higher angular resolution for tens of nearby galaxies, while simultaneously collecting more sensitive and resolved polarization data.

\section*{Acknowledgments}

We thank the KAT-7 team for supporting these observations and for their invaluable help in understanding the instrument and its calibration. We also thank David Malin for permission to use his deep image of M~83, created from data that were derived from plates taken with the UK Schmidt Telescope. G.~H. would like to thank B\"arbel Koribalski for a very useful discussion about an advanced draft of this paper. W.~J.~G.~dB was supported by the European Commission (grant FP7-PEOPLE-2012-CIG \#333939). The work of C.~C. and T.~J. is based upon research supported by the South African Research Chairs Initiative (SARChI) of the Department of Science and Technology (DST), the Square Kilometre Array South Africa (SKA SA) and the National Research Foundation (NRF). The research of D.~L., E.~E. and T.~H.~R. has been supported by SARChI \& SKA SA fellowships. L.~vZ. acknowledges support from ASTRON's Helena Kluyver Visitor Programme. The Digitized Sky Surveys were produced at the Space Telescope Science Institute under U.S. Government grant NAG W-2166. The images of these surveys are based on photographic data obtained using the Oschin Schmidt Telescope on Palomar Mountain and the UK Schmidt Telescope. The plates were processed into the present compressed digital form with the permission of these institutions. This research has made use of the NASA/IPAC Extragalactic Database (NED), which is operated by the Jet Propulsion Laboratory, California Institute of Technology, under contract with the National Aeronautics and Space Administration.

\footnotesize{
\bibliographystyle{mn2e}
\bibliography{m83kat7}
}

\bsp

\label{lastpage}

\end{document}